\documentclass[usenatbib,usegraphicx,epsfig]{aa}

\usepackage{graphicx}
\usepackage{longtable}
\usepackage{numprint}
\usepackage[multiple]{footmisc}
\usepackage{paralist}
\usepackage{comment}
\usepackage{marvosym}
\usepackage{color,soul}

\begin{document}

\newcommand{\za}{$z_{\rm abs}$}
\newcommand{\zc}{$z_{\rm cluster}$}
\newcommand{\ze}{$z_{\rm em}$}
\newcommand{\zave}{$\langle z \rangle$}
\newcommand{\nh}{$N_{\rm hit}$}
\newcommand{\vdag}{(v)^\dagger}
\newcommand{\NHI}{$N($\ion{H}{1}$)$}
\newcommand{\hi}{\ion{H}{i}}
\newcommand{\ciii}{\ion{C}{iii]}}
\newcommand{\mgii}{\ion{Mg}{ii}}
\newcommand{\mgi}{\ion{Mg}{i}}
\newcommand{\civ}{\ion{C}{iv}}
\newcommand{\siiv}{\ion{Si}{iv}}
\newcommand{\feii}{\ion{Fe}{ii}}
\newcommand{\icm}{cm$^{-2}$}
\newcommand{\kms}{km~s$^{-1}$}
\newcommand{\lya}{Ly$\alpha$\ }
\newcommand{\lyb}{Ly$\beta$\ }
\newcommand{\hb}{H$\beta$\ }
\newcommand{\nav}{$N_{\rm a}(v)$}
\newcommand{\hkpc}{$h^{-1}_{71}$ kpc}
\newcommand{\hmpc}{$h^{-1}_{71}$ Mpc}

\title{
XQ-100: A legacy survey of one hundred $3.5 \lesssim z \lesssim 4.5$ quasars
observed with VLT/XSHOOTER 
\thanks{Based on observations made with ESO Telescopes at the La Silla Paranal
  Observatory under programme ID 189.A-0424.}\thanks{The XQ-100 raw data
  and the XQ-100 Science Data Products can be found at {\tt
    http://archive.eso.org/eso/eso\_archive\_main.html} and {\tt
    http://archive.eso.org/wdb/wdb/adp/phase3\_main/form}, respectively }}
\author{S. L\'opez\inst{1}, V. D'Odorico\inst{2}, S. L. Ellison\inst{3},
  G. D. Becker\inst{4,10}, 
L. Christensen\inst{5}, 
G. Cupani\inst{2},
K. D. Denney\inst{6}, 
I. P\^aris\inst{2}, 
G. Worseck\inst{7},
T. A. M. Berg\inst{3}, 
S. Cristiani\inst{2,8}, 
M. Dessauges-Zavadsky\inst{9}, 
M. Haehnelt,\inst{10}
F. Hamann\inst{11}, 
J. Hennawi\inst{7}, 
V. Ir\v{s}i\v{c}\inst{12}, 
T.-S. Kim\inst{2},
P. L\'opez\inst{1}, 
R. Lund Saust\inst{5},
B. M\'enard\inst{13}, 
S. Perrotta\inst{14,2}, 
J. X. Prochaska\inst{15},
R. S\'anchez-Ram\'irez\inst{16,17,18} 
M. Vestergaard,\inst{5,19} 
M. Viel\inst{2,8}, 
and 
L. Wisotzki\inst{20}
}
 \institute{
Departamento de Astronom\'ia, Universidad de Chile,
  Casilla 36-D, Santiago, Chile. E-mail: {\tt
    slopez@das.uchile.cl} 
\and
INAF-Osservatorio Astronomico di Trieste, Via Tiepolo 11, I-34143 Trieste,
  Italy
\and
Department of Physics and Astronomy, University of Victoria,
  Victoria, BC V8P 1A1, Canada 
\and
Space Telescope Science Institute, 3700 San Martin Drive, Baltimore, MD 21218, USA
\and
Dark Cosmology Centre, Niels Bohr Institute,
  University of Copenhagen, Juliane Maries Vej 30, DK-2100 Copenhagen,
  Denmark
\and 
Department of Astronomy, The Ohio State University, 140
  West 18th Avenue, Columbus, OH 43210, USA 
\and 
  Max-Planck-Institut für Astronomie, Königstuhl 17, D-69117 Heidelberg,
  Germany 
\and 
INFN / National Institute for Nuclear Physics, Via Valerio 2, I-34127 Trieste, Italy
\and 
Observatoire de Genève, Université de Genève,
  51, Ch. des Maillettes, 1290, Sauverny, Switzerland 
\and 
Institute of Astronomy and Kavli Institute of Cosmology, Madingley Road, Cambridge CB3 0HA, UK
\and
Department of
  Astronomy, University of Florida, Gainesville, FL 32611-2055, USA 
\and  
International Center for Theoretical Physics (ICTP), Strada Costiera 11 - I-34151 Trieste, Italy
\and  
Department of Physics and Astronomy, Johns Hopkins
  University, 3400 North Charles Street, Baltimore, MD 21218, USA ; Alfred
  P. Sloan Fellow 
\and
International School for Advanced Studies (SISSA) via Bonomea, 265  I-34136
Trieste, Italy
\and
Department of Astronomy and Astrophysics,
UCO/Lick Observatory, University of California, 1156 High Street, Santa Cruz, CA 95064, USA 
\and  
Unidad Asociada Grupo Ciencias Planetarias (UPV/EHU, IAA-CSIC), Departamento
de F\'{\i}sica Aplicada I, E.T.S. Ingenier\'{\i}a, Universidad del Pa\'{\i}s
Vasco (UPV/EHU), Alameda de Urquijo s/n, E-48013 Bilbao, Spain 
\and
Ikerbasque, Basque Foundation for Science, Alameda de Urquijo 36-5, E-48008
Bilbao, Spain 
\and Instituto de Astrof\'{\i}sica de Andaluc\'{\i}a (IAA-CSIC),
Glorieta de la Astronom\'{\i}a s/n, E-18008, Granada, Spain 
\and  
Department of Astronomy and Steward Observatory University of Arizona, 933 N Cherry Avenue Tucson AZ 85721, USA
\and
Leibniz-Institut für Astrophysik Potsdam (AIP),
  An der Sternwarte 16, 14482, Potsdam, Germany 
}

\date{
}


\abstract{ We describe the execution and data reduction of the European
  Southern Observatory Large Programme ``Quasars and their absorption lines: a
  legacy survey of the high-redshift universe with VLT/XSHOOTER'' (hereafter
  `XQ-100'). XQ-100 has produced and made publicly available a homogeneous and
  high-quality sample of echelle spectra of $100$ quasars (QSOs) at redshifts
  $z\simeq3.5$--$4.5$ observed with full spectral coverage from $315$ to
  $2\,500$ nm at a resolving power ranging from $R\sim 4\,000$ to $7\,000$,
  depending on wavelength.  The median signal-to-noise ratios are $33$, $25$
  and $43$, as measured at rest-frame wavelengths $1\,700$, $3\,000$ and
  $3\,600$ \AA, respectively. This paper provides future users of XQ-100 data
  with the basic statistics of the survey, along with details of target
  selection, data acquisition and data reduction. The paper accompanies the
  public release of all data products, including 100 reduced spectra. XQ-100
  is the largest spectroscopic survey to date of high-redshift QSOs with
  simultaneous rest-frame UV/optical coverage, and as such enables a wide
  range of extragalactic research, from cosmology and galaxy evolution to AGN
  astrophysics.  }


\authorrunning{S. L\'opez et al.}
\titlerunning{XQ-100}

\maketitle

\section{Introduction}
\label{sect_intro}

%

In the era of massive quasar (QSO) surveys, already encompassing hundreds of
thousands of confirmed sources~\citep[e.g.,][]{paris2014,flesch2015}, there is
a relative shortage of follow-up echelle quality spectroscopy.  Moderate to
high resolving power ($R \approx 5\,000$--$40\,000$) and wide spectral
coverage are key to many absorption line diagnostics that probe the
interplay between galaxies and the intergalactic medium (IGM) at all
redshifts. However, such observations are time consuming and require large
telescopes, and even more so for high redshift QSOs which tend to be
faint. Another challenge for QSO absorption line science is that as the
redshift increases, more of the rest-frame UV and optical transitions become
shifted into the hard-going near-infrared (NIR; $1\mu{\rm m} \la \lambda \la
2.5 \mu $m).  Presently, public archives contain echelle spectra of
roughly a few thousand unique QSOs, of which just a small fraction has NIR
coverage. In addition, these data arise primarily from the cumulative effort
of single (and heterogenous) observing programs, so one would expect such
databases to be inhomogeneous in nature and suffer from selection biases by
construction\citep{brunner2002,djorgovski2005}. Thus, new homogeneous and
statistically significant echelle data sets are always welcome with as wide a
range of uses as possible. In this paper we present ``XQ-100'', a new legacy
survey of 100 QSOs at emission redshifts $z_{\rm em}\simeq 3.5$--$4.5$
observed with full optical and NIR coverage using the echelle spectrograph
XSHOOTER~\citep{Vernet2011} 
 on the European Southern Observatory (ESO) Very Large Telescope
(VLT). The context and the scientific motivation of the survey are as follows.

The largest QSO echelle samples in the optical come from
Keck/HIRES~\citep[``KODIAQ'' database; ][]{omeara2015} and VLT/UVES (ESO UVES
public archive) each providing between $300$ and $400$ QSO spectra with
$R\approx 40\,000$. At moderate resolving power, $R\approx 10\,000$, Keck/ESI
has observed around a thousand QSOs (John O'Meara, private communication) and
a search in the VLT/XSHOOTER public archive reveals spectra of almost $300$
sources to date.  Other large optical facilities with echelle capabilities,
such as Subaru or Magellan, have either acquired a smaller data volume or do
not manage public archives.  In addition to ``smaller'' programs ($\la 10$
targets), these data sets, public or not, have been fed over the years by a
few dedicated QSO surveys~\citep[e.g.,][]{bergeron2004} aimed at a variety of
astrophysical probes of galaxy evolution and cosmology: metals in
damped \lya\ systems~\citep[DLAs; e.g.,][]{lu1996,
  prochaska2003,ledoux2003,rafelski2012} and in the
IGM~\citep[e.g.,][]{aguirre2004,songaila2005,scannapieco2006,dodorico2010};
light elements in Lyman-limit Systems~\citep[e.g.,][]{ kirkman2003}; DLA
galaxies~\citep[e.g.,][]{peroux2011,noterdaeme2012a,zafar2013}; low and high-z
circum-galactic medium~\citep[e.g.,][]{chen2010,rudie2012}; thermal state of
the IGM~\citep[e.g.,][]{schaye2000,kim2002};
reionization~\citep[e.g.,][]{becker2007,becker2012,becker2015}; matter power
spectrum~\citep[e.g.,][]{croft2002,viel2004,viel2009,viel2013}; and
fundamental constants~\citep[e.g.,][]{murphy2003,srianand2004,molaro2013}.

In the NIR, the largest QSO spectroscopic survey so far has been conducted
using the FIRE IR spectrograph at Magellan~\citep{Matejek2012}.  Focused on
the incidence of \mgii\ at $z\approx2$--$5$, this survey comprises NIR
observations of around 50 high-$z$ QSOs at $R\approx 6\,000$ and median
signal-to-noise ratio, S/N $=13$. Other surveys at moderate to high resolution
have focused on the \civ\ mass density at $z>4$ using
Magellan/FIRE~\citep{Simcoe2011},
Keck/NIRSPEC~\citep{becker2009,ryan2009,becker2012}, or
VLT/XSHOOTER~\citep{dodorico2013}, albeit comprising only a handful of
sightlines, given the paucity of very high-$z$ QSOs.

Near-IR spectroscopy is also needed to study the rest-frame optical emission
lines of high-$z$ QSOs, which constrain broad-line region metallicities and
black hole masses; however, in this case spectral coverage is more important
than resolution.  For instance, surveys have used 
VLT/ISAAC~\citep{sulenticetal06,sulenticetal04,marziani2009},
NTT/SofI~\citep{dietrich2002,dietrich2009}, or Keck/NIRSPEC and
Blanco/OSIRIS~\citep{dietrich2003}. There are also samples at higher
resolution obtained with Gemini/GNIRS~\citep{jiang2007}, or
VLT/XSHOOTER~\citep{ho2012,derosa2014}. The largest samples have been acquired
using Palomar Hale 200-inch/TripleSpec~\citep[][32 QSOs at
  $3.2<z<3.9$]{zuo2015} and, at lower redshifts, VLT/XSHOOTER~\citep[][30 QSOs
  at $z\approx1.5$]{capellupo2015}.

The present XQ-100 survey builds on observations made with VLT/XSHOOTER within
the ESO Large Programme entitled ``Quasars and their absorption lines: a
legacy survey of the high redshift universe with X-shooter'' (PI S. L\'opez;
$100$ hours of Chilean time).  XSHOOTER provides complete coverage from the
atmospheric cutoff to the NIR in one integration at $R\approx
6\,000$--$9\,000$, depending on wavelength.  The full spectral coverage, along
with a well-defined target selection and the high S/N 
achieved (median S/N $=30$), clearly make XQ-100 a unique data set to study the
rest-frame UV/optical of high-$z$ QSOs in a single, homogeneous, and
statistically significant sample. Our program was based on the following
scientific themes:

\begin{enumerate}

\item {\it Galaxies in absorption:} determining the cosmic density of neutral
  gas in DLAs, the main reservoirs of neutral gas
  in the Universe~\citep[e.g.,][]{wolfe2005, prochaska2009a,noterdaeme2012b} 
  at $z>3.5$~\citep{sanchez2015}; studying individual DLA abundances at
  $2.0\la z\la 4.5$~\citep{berg2016}; constraining the \mgii\ incidence
  $(dN/dz)_\mgii$ at $z>2.5$ with $\sim 2$--$3$ times better sensitivity and
  $\sim 2$ times longer redshift path than the sample by~\citet{Matejek2012}
  to test predictions from the cosmic star formation
  rate~\citep{zhu2013,2011MNRAS.417..801M}.

\item {\it Intergalactic-Medium science:} measuring the cosmic opacity at the
  Lyman limit~\citep{prochaska2009,worseck2014} and providing an independent
  census of Lyman-limit systems~\citep[LLS;][]{prochaska2010,songaila2010} at
  $z\simeq1.5$--$4.5$; constraining the UV background via the proximity
  effect~\citep[e.g.,][]{dodorico2008,dallaglio2008,calverley2011}.

\item {\it Active-Galactic-Nuclei science:} making the first $z> 3.5$ accurate
  measurements of black hole masses using the rest-frame UV emission lines
    of \civ$\lambda 1549$ and \mgii$\lambda 2800$ and the rest-frame optical
  \hb\ line \citep[from line widths and continuum luminosities;
      e.g.,][]{vestergaard2006,vestergaard2009}; examining broad-line region
    metallicity estimates (from emission line ratios; e.g., Hamann \& Ferland
    1999; Hamann et al. 2002) and their relationship with
    other QSO properties, including, but not limited to, luminosity and black
    hole mass; using associated absorption lines to study the co-evolution of
  galaxies and black holes by measuring metallicities in the
  interstellar-medium of the QSO host
  galaxies~\citep[][]{perrotta2016,dodorico2004}; studying the broad
    QSO-driven outflow 
    absorption lines that are found serendipitously in the spectra.

\item 
{\it Cosmology:} measuring the matter power spectrum with the Ly$\alpha$
forest~\citep{croft1998} at high
redshift~\citep[e.g.,][]{viel2009,palanque2013}, including an independent
measurement of cosmological parameters with a joint analysis of these and the
Planck publicly released data~\citep{irsic2016}.

\end{enumerate}

The sample size of 100 QSOs was defined by the objectives of these science
goals. The choice of emission redshifts was determined by the absorption line
searches: $z \ga 3.5$ means that every QSO contributes a redshift path of at
least $0.5$ for $(dN/dz)_\mgii$ in the NIR, while $z \la 4.5$ avoids excessive
line crowding in the Ly$\alpha$ forest.  Clearly, a combination of the
  factors: well-defined target selection, echelle resolution, high S/N, and
  full wavelength coverage all represent a benefit to the above science
  goals.

XQ-100 was designed as a legacy survey and this paper accompanies the public
release of all data products, including a uniform sample of 100 reduced
XSHOOTER spectra (available at {\tt http://archive.eso.org}).  We note that this
data volume increases the XSHOOTER QSO archive by $\approx 30 \%$.

The following sections provide an in-depth description of the survey, along
with its basic statistics. A description of our target selection and the
observations can be found in Section \S~\ref{sect_target_and_observations}; details of
the data reduction, along with a comparison between our own custom pipeline
and the one provided by ESO are given in \S~\ref{section_data_reduction};
details of data post-processing (telluric corrections and continuum fits) are
given in \S~\ref{section_post}; and, finally, a description of the publicly
released data products is given in \S~\ref{section_products}. For a technical
description of the instrument, we refer the reader to~\citet{Vernet2011} and
to the online XSHOOTER documentation.\footnote{\tt
  http://www.eso.org/sci/facilities/paranal/
  instruments/xshooter/doc.html}\footnote{\tt
  http://www.eso.org/observing/dfo/quality/
  XSHOOTER/qc/problems/problems$\_$xshooter.html}

\section[]{Target selection and observations}
\label{sect_target_and_observations}

\subsection{Target selection}

XQ-100 targets were selected initially from the NASA/IPAC Extragalactic
Database (NED) to have emission redshifts $z > 3.5$ and declinations $\delta <
+15$ degrees.  To fill some right-ascension gaps lacking bright $z > 3.5$
targets, twelve additional targets with $+15 < \delta < +30$ were selected
from literature sources.  Then the Sloan Digital Sky Survey Data Release 7
database~\citep[SDSS DR7;][]{schneider2010} was screened with the further
criterion of having SDSS magnitude $r < +20$.  Finally, these candidates were
cross-correlated with the Automate Plate Machine (APM) catalog\footnote{\tt
  http://www.ast.cam.ac.uk/$\sim$mike/apmcat/} to obtain uniform magnitudes in
a single pass-band ($R$), which we also use throughout the present paper.
Our primary selection is thus biased toward bright sources; however, as
  explained below, we made our best effort to minimize biases affecting the
  absorption line statistics.

We avoided targets with known broad absorption line features, and targets with
an intrinsic color selection bias from the SDSS.  The SDSS color selection is
biased at the lower redshift end of our survey~\citep[$z<3.6$,
  see][]{Worseck2011}.  Here, we required SDSS QSOs to be radio-selected or
previously discovered with other techniques such as slitless
spectroscopy. Without these precautions, our goal of obtaining a truly blind
and unbiased target selection would have been undermined, despite the
relatively small number of targets impacted.  For example the SDSS color bias
would result in (1) underestimates of the mean free
path~\citep{prochaska2009}; (2) overestimates of the DLA --and also the LLS--
incidence~\citep{prochaska2010}; (3) a higher metal $dN/dz$ due to the higher
incidence of LLSs and partial LLSs; (4) a higher fraction of proximate LLSs
that affect proximity effect studies; and (5) potentially a slight bias in the
mean QSO spectral energy distribution towards red QSOs~\citep{Worseck2011}. We
should also note that although earlier color survey designs (Palomar
Spectroscopic Survey, APM BR, APM BRI)
considered color selection effects at the low-z
end~\citep{irwin1991,storrie1994}, these were never well quantified. Thus,
follow-up on color-selected QSOs close to the stellar locus should be done
with care (or avoided altogether), as the sightlines are potentially biased in
their LLS statistics.

During program execution we replaced four targets in our original list that
had been observed by~\citet{Matejek2012} using Magellan/FIRE; however, we
intentionally observed three other FIRE targets in order to have a reference
in characterizing absorption line detection limits: 
J1020+0922 at $z=3.640$, J1110+0244 $z=4.146$, and J1621-0042 at
$z=3.711$.

Our final sample, taking into account the various selections described above
and also considering the relative paucity of high redshift QSOs, has emission
redshifts ranging from $3.508$ to $4.716$.  Since the most distant QSO in our
sample is the only target with $z_{\rm em} > 4.5$, for simplicity we refer to
the redshift range of the survey as $z_{\rm em}\simeq 3.5$--$4.5$ throughout
this paper.

\begin{figure}
\centering
\includegraphics[width=90mm,angle=0,clip]{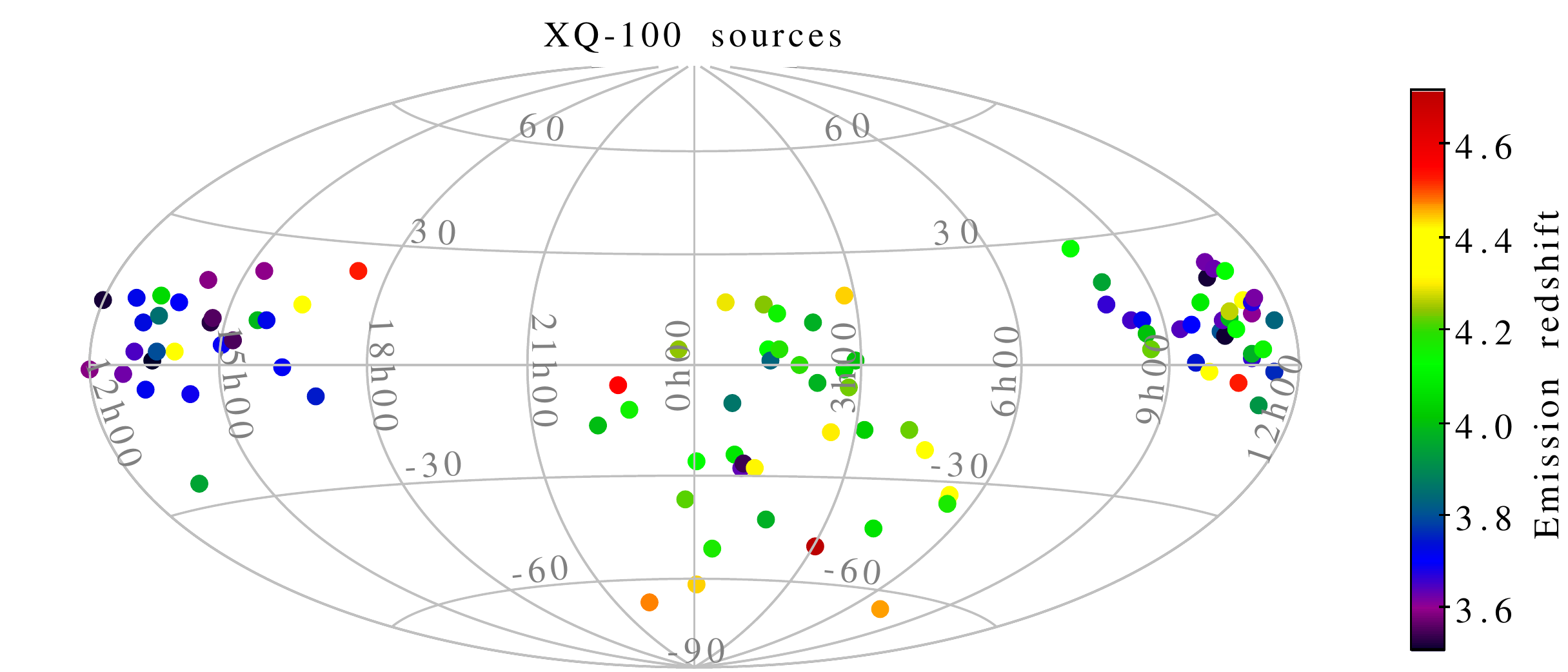}
\caption{Sky distribution of XQ-100 sources. The color scale indicates
  emission redshifts. \label{fig_sky}}
\end{figure}

\begin{figure}[b]
\includegraphics[width=84mm,clip]{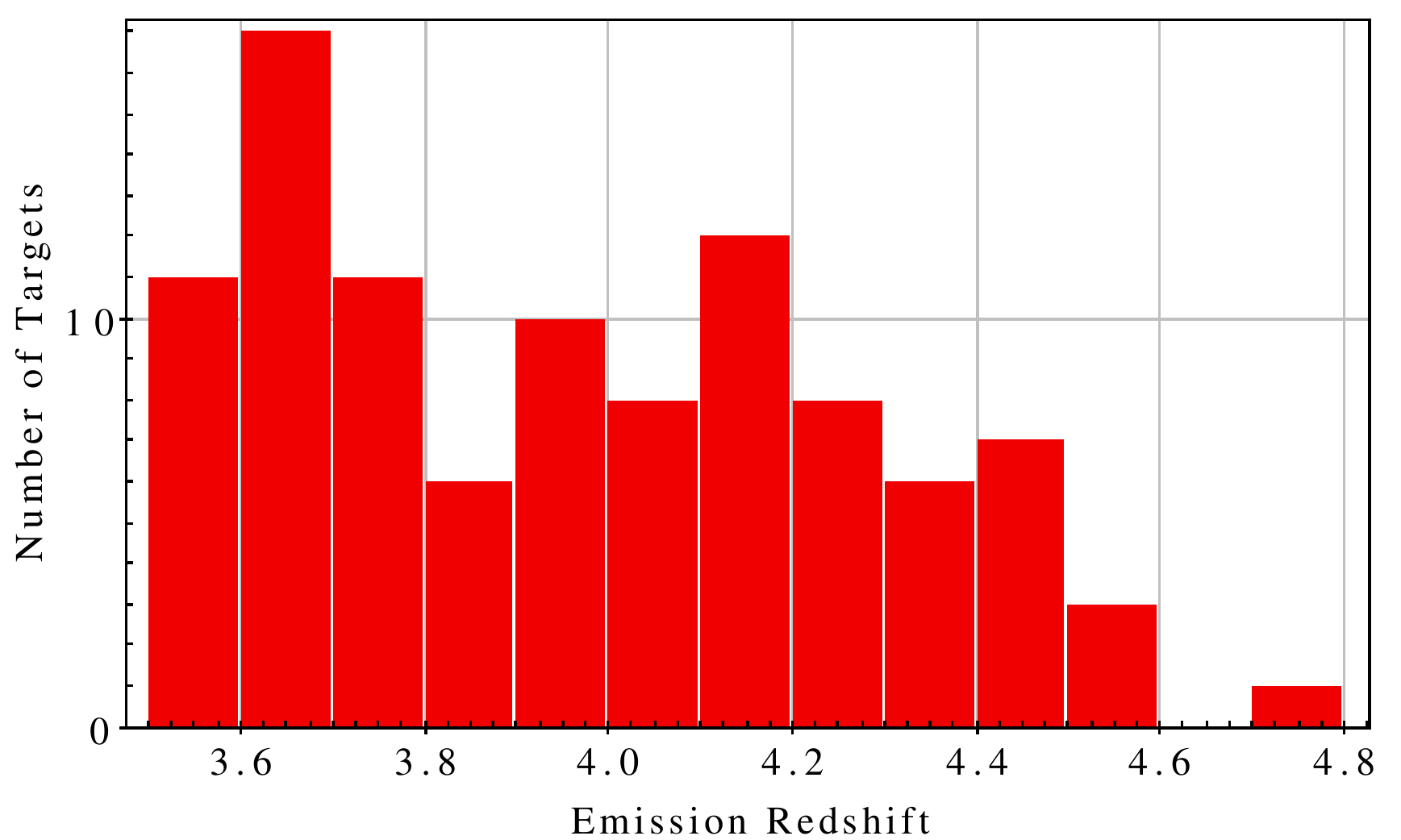}
\caption{XQ-100 emission redshifts. \label{fig_z}}  
\end{figure}

\begin{figure}[b]
\includegraphics[width=84mm,clip]{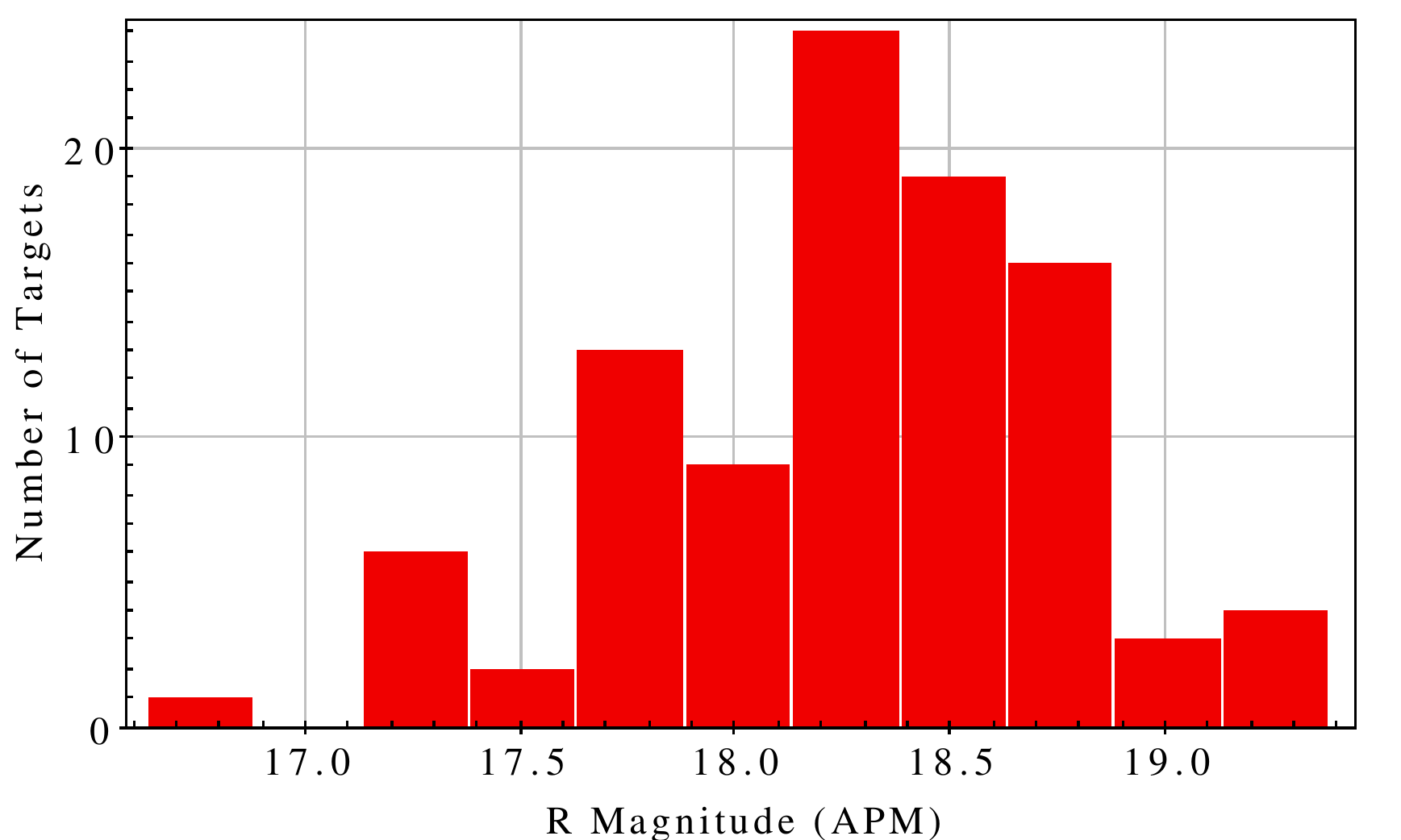}
\caption{XQ-100 $R$-magnitudes (APM). \label{fig_histo}}  
\end{figure}

Figure~\ref{fig_sky} shows the sky distribution of the
observed XQ-100 sample. A color scale depicts emission redshifts.
Figures~\ref{fig_z} and~\ref{fig_histo} show the final distribution of QSO
emission redshifts and $R$-magnitudes, respectively.

The full target list is provided in Table~\ref{table_targets} of the Appendix,
along with basic target properties (see Section~\ref{section_data_reduction}).
A full catalog with all observed target properties (listed in
Table~\ref{table_parameters}) is provided online along with the data at {\tt
  http://archive.eso.org/eso/eso\_archive\_main.html}.

\subsection{Observations}
\label{sect_observations}

The observations were carried out in ``service mode'' between April 1, 2012, and
March 26, 2014. During this time XSHOOTER was mounted on unit 2 of the
VLT. Service mode allows the user to define the Observation Blocks
(OBs), which contain the instrument setup and are carried out by the
observatory under the required weather conditions.

Table~\ref{table_weather} summarizes the requested conditions of XQ-100. The
airmass constraint was set according to each target's declination such that
the target was observable above the set constraint for at least 2 hours. The
requested constraints on sky brightness were fraction of lunar illumination
$<0.5$ and minimum moon distance 45 degrees. The targets were split into
two samples, brighter and fainter than magnitude $R_{\rm APM}=18.0$.  The
seeing constraint was set to $1.0$\arcsec\ for the bright sample and
$0.8$\arcsec\ for the faint sample. ESO Large Programmes are granted high
priority status, which means that observations out of specifications are
repeated and eventually carried over to the following semester until the
constraints are met (to within $\approx 10$ \%).  In our case 13 targets were
observed more than once because of interrupted OBs or because of ADC issues
(\S~\ref{sect_ADC}).\footnote{The number of OB executions is listed in column
  5 of Table~\ref{table_parameters}} As a consequence of this process, 88
XQ-100 targets were observed within specifications, and 12 almost within
specifications (i.e., the constraints were worse by $\lesssim 10$ \%).

\begin{table} 
 \centering
 \begin{minipage}{140mm}
  \caption{Requested observing conditions\label{table_weather}}
  \begin{tabular}{lr}
  \hline
 Seeing  & $1.0$\arcsec\ (bright), $0.8$\arcsec\ (faint) \\
 Sky transparency  & Clear \\
         & $\delta > +20$: $< 1.6$\\
 Airmass & $+10<\delta <20$: $< 1.5$\\
         & $0<\delta < +10$: $< 1.4$\\
         & $\delta <0$: $< 1.3$\\
 \% of lunar illumination     & 50\% \\
 Moon distance & 45 degrees\\
\hline
\end{tabular}
\end{minipage}
\end{table}

\begin{table*}
 \centering
 \begin{minipage}{160mm}
  \caption{Instrument setup\label{table_observations}}
  \begin{tabular}{cccccccc}

  \hline

   Arm & Wavelength range & Slit width  & Resolving power &\multicolumn{2}{l}{Num. of exposures} &
   \multicolumn{2}{l}{Integration time (s)}\\ 
       &  [nm]    &   (\arcsec)            & $\lambda/\Delta\lambda$ & bright & faint & bright &
   faint\\  

 \hline

 UVB  & 315--560       & 1.0 & 4\,350 & 2 & 4 & 890 & 880 \\ 
 VIS  & 540--1\,020    & 0.9 & 7\,450 & 2 & 4 & 840 & 830 \\ 
 NIR  & 1\,000--2\,480$^a$ & 0.9 & 5\,300 & 2 & 4 & 900 & 900 \\ 

\hline

\end{tabular}
$^a${\footnotesize 1\,000--1\,800 nm when the $K$-band filter was   used; 
    see~\S~\ref{sect_observations}.}  
\end{minipage}
\end{table*}

Table~\ref{table_observations} summarizes the instrument setup. XSHOOTER has
three spectroscopic arms, UVB, VIS and NIR, each with its own set of shutter,
slit mask, cross-dispersive element, and detector. In order to obtain
signal-to-noise ratios, that are as uniform as possible, XQ-100 integration
times varied across the samples and also across the three spectroscopic
arms. The bright sample had two integrations, each with $T_{\rm exp }=890$s in
UVB, $T_{\rm exp }=840$s in VIS and $T_{\rm exp }=900$s in the NIR. The faint
sample had four exposures, each with $T_{\rm exp }=880$s in the UVB, $T_{\rm exp
}=830$s in the VIS, and $T_{\rm exp }=900$s in the NIR.  These conditions
defined two classes of OBs, which -- including acquisition -- had a total of 
$39$ and $70$ minutes duration, respectively. In order to optimize the
sky-subtraction in the NIR, the exposures were nodded along the slit by $\pm
2.5$\arcsec\ from the slit center.

The adopted slit widths were 1.0\arcsec\ in the UVB and 0.9\arcsec\ in the
VIS and NIR, to match the requested seeing and account for its wavelength
dependence.  These slit widths provide a nominal resolving power of $4\,350$,
$7\,450$, and $5\,300$, respectively.  The slit position was always set along
the parallactic angle, except for five targets for which it was necessary to
avoid contamination of a nearby bright object in the slit; these cases are
relevant to a problem with the atmospheric dispersion corrector system (see
next Section). Target acquisition was done in the $R$ filter. The UVB and VIS
were binned by a factor 2 in the dispersion direction.

For emission redshifts $z > 4$, the [OIII]$\lambda$5007 emission line lies out
of the $K$-band.  For $4.0 \la z \la 4.5$, [OII]$\lambda$3727 falls in the gap
between the $H$- and $K$-bands.  Therefore, the 53 XQ-100 sources having $z>4$
were observed using a $K$-band blocking filter that lowers the sky background
where scattered light from the $K$-band affects primarily the
$J$-band~\citep{Vernet2011}.  No blocking filter was used for $z<4$ sources
(47) in order to include [OIII]$\lambda$5007 in the wavelength range. We note 
that \mgii$\lambda\lambda$2796,2803 is always in the wavelength range. See
Fig.~\ref{fig_spec} for an example of a spectrum presenting the
above-mentioned emission lines.

For each exposure, the standard calibration plan of the observatory was used
to observe a hot star for telluric corrections. This plan foresees the
observation of a telluric standard within 2 hours and 0.2 airmasses of each
science observation (but see~\S~\ref{section_telluric}).

\subsubsection{ADC issues}
\label{sect_ADC}

In March 2012 ESO reported that the atmospheric dispersion correctors (ADCs)
of the UVB and VIS arms started to fail occasionally, leading to possible
wavelength-dependent slit losses, potentially worse than if no ADCs were
used. In August 2012 the ADCs were disabled for the rest of the observations
(at the time of writing the causes of these failures are being investigated).

\begin{figure}
\includegraphics[width=88mm,clip]{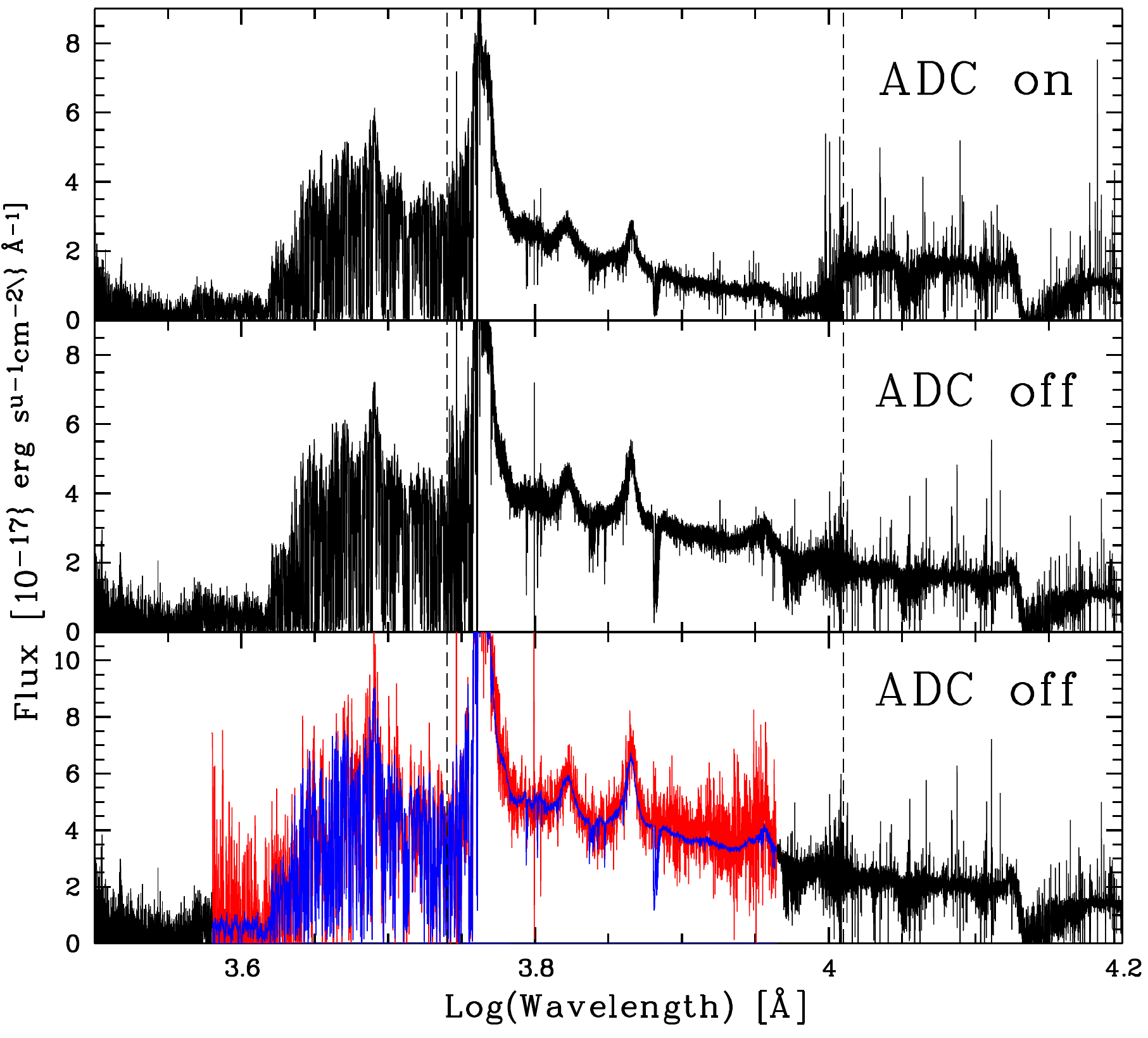}
\caption{XQ-100 spectra of the same QSO,
  J1126$-$0124, taken with the faulty ADCs in April 2012 (top panel) and
  repeated with the disabled ADCs in February 2014 (middle panel). Both
  observations were executed at a similar airmass of $\approx 1.3$ at the
  parallactic angle.  The dashed lines indicate the boundaries of the XSHOOTER
  arms. The match is better between the VIS and NIR arms in the middle
  panel. The bottom panel shows the same February 2014 XQ-100 spectrum but
  smoothed and rebinned to SDSS resolution (blue line), and rescaled by a
  factor of 1.3 to match the corresponding SDSS spectrum (overlaid in
  red). The good match suggests that slit losses in the XQ-100 data are
  roughly achromatic.
\label{fig_ADC}}
\end{figure}

By August 2012, around 30\% of the XQ-100 observations had been
executed. After checking our spectra carefully, we noticed the ADC problem had
possibly affected 12 of the spectra, which showed an unusually large flux
mismatch between the arms (see example in top panel of Fig.~\ref{fig_ADC},
which is explained below). The reason for such a mismatch was probably that
these targets had been observed at a high enough airmass for a malfunctioning
ADC to lead to strong chromatic slit losses.

Five out of these 12 OBs were executed a second time with the disabled ADCs
and using the parallactic angle.  The improvement was evident.  The two upper
panels of Fig.~\ref{fig_ADC} show XQ-100 spectra of the same OB executed
before and after the ADC disabling.  We note the effect of the faulty
ADCs on the flux levels and slope in the UVB and VIS arms only (top panel),
while the NIR arm is not affected, which is expected since this arm does not
use an ADC. Conversely, without the ADCs (middle panel) the flux levels have a
better match between the arms (spectra were taken at the parallactic angle
always).  The bottom panel of Fig.~\ref{fig_ADC} shows the XQ-100 spectrum
from the middle panel but smoothed and rebinned to SDSS resolution (blue
line), and rescaled by a factor of 1.3 to match the corresponding SDSS
spectrum (overlaid in red). The good match across wavelengths suggests that
slit losses, at least in the SDSS spectral region, are roughly achromatic in
the XQ-100 spectra.

Since the accuracy of flux calibrations is unimportant for many of the science
applications described in the introduction and an extra exposure might be
helpful to increase the S/N, we provide reduced spectra of both observations
in these 13 cases and flag them in our database (see 
Section~\ref{section_products}).

The remaining observations in the queue proceeded without the ADCs but making
sure that the parallactic angle and the lowest possible airmass was chosen. 

\section{Data reduction}
\label{section_data_reduction}

\begin{figure}
\includegraphics[width=92mm,clip]{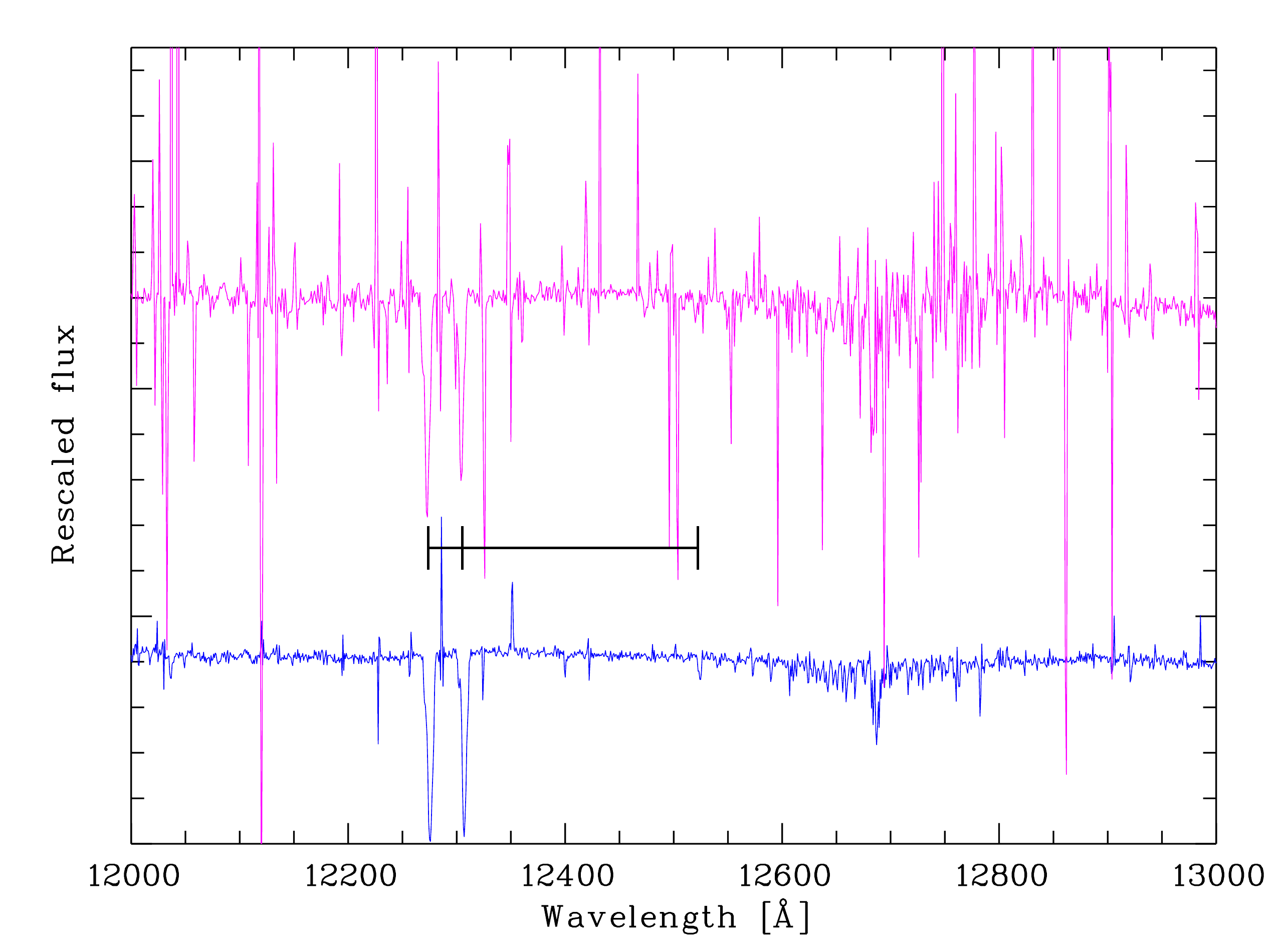}
\caption{Portion of the NIR spectrum of QSO J0003-2603 at $z=4.125$ reduced
  using the XSHOOTER/Reflex pipeline, version 2.5.2 (top) and our own {\sc
    idl} pipeline (bottom). The tickmarks in between spectra indicate 
  \mgii\ and \mgi\ absorption lines in a DLA at $z=3.390$. The 
  bottom spectrum is much less affected by the
  residuals. \label{fig_pipeline}}
\end{figure}

Extraction of NIR spectra can prove a non-trivial task owing to the high
sky-background levels.  ESO provides a pipeline to reduce XSHOOTER data, which
we have tested. However, in doing so, we noticed that the reduced spectra show
systematically large and frequent sky-subtraction residuals in the
NIR. Consequently, we opted to implement our own custom pipeline and to reduce
XQ-100 data using scripts written in {\sc idl} by one of us
(GDB). Figure~\ref{fig_pipeline} shows an example that highlights the
differences between the two pipelines in the NIR.  Overall, despite some
unavoidable residuals, the {\sc idl} pipeline seems to be more effective than
the ESO version available by mid-2014.  In the following two sections we
describe our pipeline and then provide a qualitative comparison with the ESO
version.

\subsection{Custom pipeline}

The overall reduction strategy is based on the techniques of
\citet{kelson2003}, where operations are performed on the un-rectified 2D
frames. To achieve this, we generated 2D arrays of slit position
and wavelength that served as the coordinate grid for sky modeling and
1D spectrum extraction.  A fiducial set of coordinate arrays for
each arm was registered to individual science frames using the measured
positions of sky and/or arc lines.

Individual frames were bias subtracted (or dark subtracted in the case of the
NIR arm) and flat-fielded.  The sky emission in each order was then modeled
using a b-spline and subtracted.  To avoid adding significant extra noise in
the NIR arm, composite dark frames were generated from multiple (typically
$\sim10$) dark exposures with matching integration times.  This approach was
found to remove the fixed pattern noise in the NIR to the extent that the sky
emission could generally be well modeled in each exposure independently,
without subtracting a nodded frame, thus avoiding a factor $\sqrt{2}$ penalty
in the background noise.  The exception to this was the reddest order
(2\,270-2\,480 nm), which is problematic because it is vignetted by a baffle
designed to mask stray light (see footnote 2 in~\S~\ref{sect_intro}) This
order was therefore nod-subtracted, and the residual sky emission modeled
using a b-spline.

Following sky subtraction, the counts in the 2D frames were flux
calibrated using response curves generated from observations of
spectro-photometric standard stars.  Standards observed close in time to the
science observations were generally used.  For a limited number of objects,
however, the temporally closest star was not optimal and unexpected features
were observed in the flux-calibrated spectra. In these cases, a fiducial
response curve was used to produce an additional flux-calibrated spectrum.

A single 1D spectrum was then extracted simultaneously from across
all orders and all exposures of a given object (in a single arm).  Extraction
was performed on the non-rectified frames to avoid multiple rebinnings and to
keep the error correlation across adjacent pixels to a minimum.  The number of
exposures for each object ranges between 2 and 12, depending on the number of
scheduled exposures (two to four) and on the number of times a given
OB was executed (typically one, but two or three in cases of
interrupted execution and ADC issues).  When observations were spread across
several nights, a separate 1D spectrum was extracted for each
night. 

The one-dimensional spectra were binned using a fixed velocity step. This is
the only rebinning involved in the reduction procedure.  Wavelength bins for
the three arms (UVB: $20$ km s$^{-1}$; VIS: $11$ km s$^{-1}$; NIR: $19$ km
s$^{-1}$) were chosen to provide roughly 3 pixels per FWHM, taking the nominal
XSHOOTER resolving power for the adopted slits (Table~\ref{table_observations}).
The whole (gap-less) wavelength range is 315 to $1\,800$ nm for spectra taken
with the $K$-band blocking filter, and $315$ to $2\,480$ nm for other
spectra. Wavelengths were corrected to the vacuum-heliocentric system. When
multiple exposures of a single object existed, they were co-added (with the
exception of exposures taken with the faulty ADC, which were not included in
the co-added spectrum). The stacking was done arm by arm; no attempt was made
to merge the arms at this stage, although we do provide joint spectra in the
public release (\S~\ref{section_products}). In the following, we call these
reduced data ``raw'' to distinguish them from the post-processed data (described
in Sections~\ref{section_post} and~\ref{section_products}).

\begin{figure}
\includegraphics[width=100mm,clip]{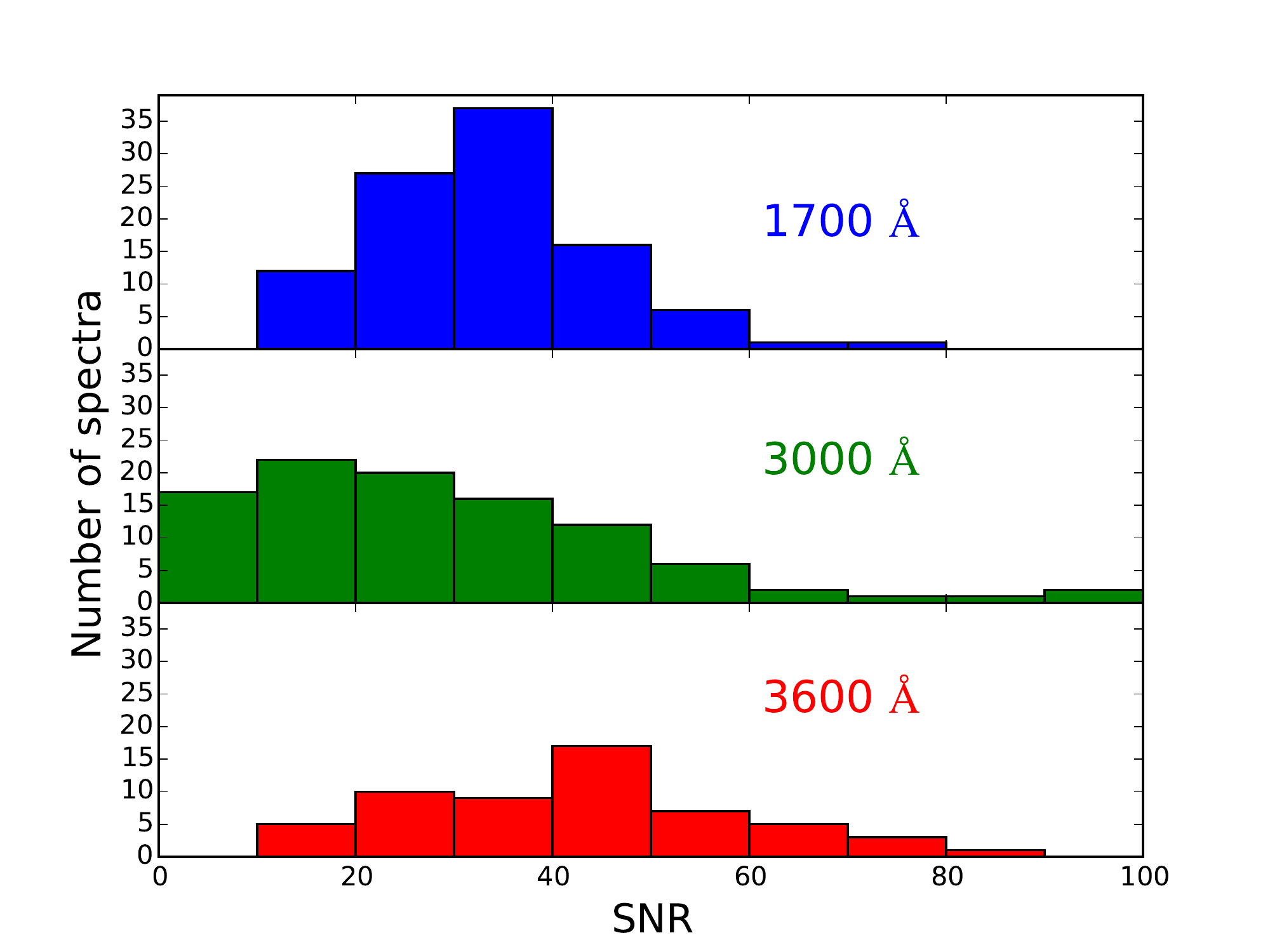}
\caption{Distribution of pixel signal-to-noise ratios of the co-added
  spectra at three different rest-frame wavelengths: $1\,700$, $3\,000$, and
  $3\,600$ \AA.  The S/N is computed in a window centered at those wavelengths
  and spanning $\pm 10$ \AA. \label{fig_snr}. The $3\,600$ \AA{}   
    histogram has fewer elements because not all spectra cover that wavelength
   (see \S~\ref{sect_observations}). }
\end{figure}

Figure~\ref{fig_snr} shows the distribution of S/N (per pixel) at three
different rest-frame wavelengths: $1\,700$ \AA\ (representative of the VIS
spectra), $3\,000$ \AA\ (NIR spectra of high-$z$ sources), and $3\,600$
\AA\ (NIR spectra of low-$z$ sources).  The respective median
  signal-to-noise ratios are $33$, $25$ and $43$, as measured in a $\pm 10$
  \AA{} window at those rest-frame wavelengths. These values are consistent
with the predictions of the XSHOOTER Exposure Time Calculator, which motivated
the setup adopted for the OBs.

Figures~\ref{fig_all} to~\ref{fig_all_last} in Appendix B show all reduced
spectra and Fig.~\ref{fig_spec} shows an example with an expanded wavelength
scale.

\subsection{Accuracy of the flux calibration}
\label{sec:eso_flux}

Comparison with SDSS spectra (expected to have little aperture loss given the
$3\arcsec$ fibers of the spectrograph) shows a systematic underestimation of
the flux on the XSHOOTER part
($\left\langle\frac{F_\mathrm{XSH}}{F_\mathrm{SDSS}}\right\rangle\sim 0.77$),
mainly due to slit losses induced by the narrow slits used.  As shown in the
bottom panel of Fig.~\ref{fig_ADC}, these slit losses appear to be roughly
achromatic.

In some cases the flux values in adjacent arms (especially VIS and NIR) do not
match exactly and a gap is observed.  In general we expect a mild mismatch
which probably depends on seeing, since slit widths are different in each arm
and the standard stars, used for flux calibration, are all taken with a
5\arcsec\ slit. 
However, in six spectra a large mismatch is observed ($\approx 30$\% across
arms), which cannot be attributed to slit losses only. These spectra are:
J0113-2803, J1013$+$40650, J1524$+$2123, J1552$+$1005, J1621$-$0042, and
J1723$+$2243. For these particular cases, three possible causes were
identified: (1) the ADC issue (\S~\ref{sect_ADC}); (2) a sudden interruption
in the OB execution, which produced UVB and VIS frames with shorter
integration time (when this happens, NIR frames are automatically discarded);
or (3) problems with flat-fielding. Since an ad hoc treatment of individual
targets was beyond the scope of this release and would have compromised the
consistency of the reduction process, we decided not to undertake any further
action in this direction.

Thus, flux calibration of XQ-100 spectra should
not be taken as absolute.  The spectral shape is correctly reconstructed and
the flux values can be taken as order-of-magnitude estimates, but users of the
public data release may want to refer to photometry when an accurate flux
measurement is needed.

\begin{figure*}
\centering
\includegraphics[width=170mm,clip]{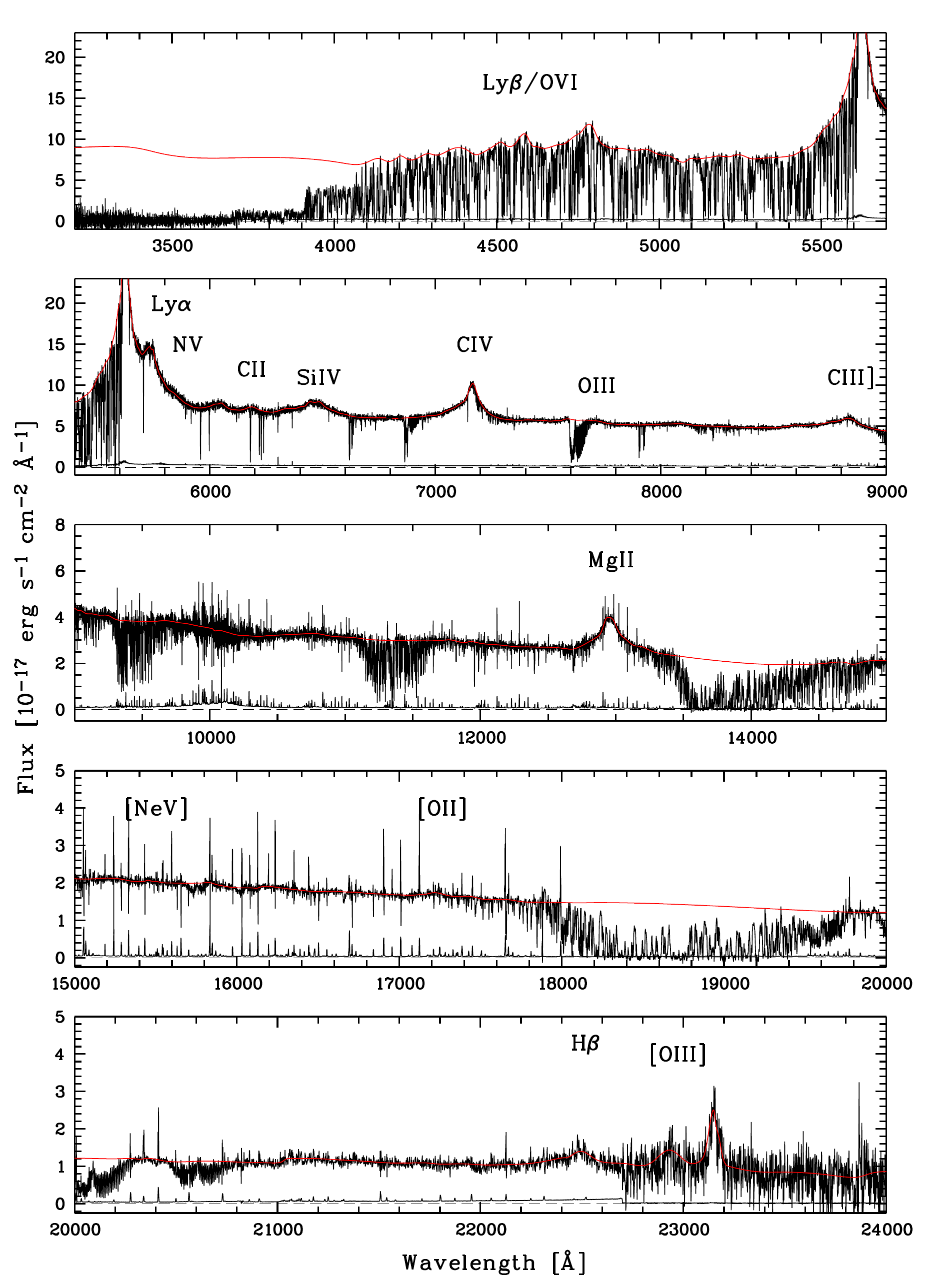}
\caption{XQ-100 spectrum of QSO J1117+1311 at $z=3.622$, a representative case
  of the whole sample in terms of S/N.  The flux is not corrected for telluric
  absorption (\S~\ref{section_telluric}) or rebinned for display purposes.
  Some emission lines are marked.  The red line depicts 
a manually placed continuum made of cubic splines
  (see~\S~\ref{section_continuum} for details).  The complete set of  
  XQ-100 spectra is shown in Figs.~\ref{fig_all} to~\ref{fig_all_last}.
\label{fig_spec}}  
\end{figure*}

\subsection{Comparison with ESO pipeline}
\label{sec:eso_idl}

The ESO pipeline is run through an environment called
Reflex~\citep{freudling2013}, which allows the user to organize the scientific
and calibration files and to execute the pipeline in an interactive and
graphical fashion.  A qualitative comparison between version 2.5.2 of the ESO
pipeline and our custom pipeline is as follows:

\begin{itemize}
\item Wavelength calibration: The ESO pipeline performs a two-step wavelength
  calibration of raw spectra, using arc lamp frames. In the first step, the
  positions of the order edges and arc lines are predicted from a physical
  model of the instrument. In the second step, a 2D mapping from the detector
  space to the $(\lambda,s)$ space is computed, where $s$ is the position of
  the pixel along the slit. This mapping is used to produce the final 2D
  rectified spectrum. Conversely, our custom-built {\sc idl} package starts with
  2D $\lambda$ and $s$ coordinate frames that have been carefully
  calibrated for a single reference exposure, and then shifts these frames to
  match other exposures using the measured positions of sky (VIS, NIR) or arc
  (UVB) lines.  As a consequence, the cascade for the {\sc idl} package is
  simpler and the overall execution time (data retrieval+processing) is
  generally shorter.

\item Sky subtraction: Both tools implement the \citet{kelson2003} algorithm
  for optimal sky subtraction. For reasons that remain unclear, the {\sc idl}
  package provides much better results than the ESO pipeline. Residuals of
  sky-line subtraction in the NIR arm are consistently higher in spectra
  obtained with the ESO pipeline, as seen in Fig.~\ref{fig_pipeline}.

\item Object tracing: In the ESO pipeline, the position of the object is
  extracted from the 2D rectified (i.e. rebinned) spectrum; in MANUAL mode,
  the position of the centroid and the trace width are both set constant. The
  {\sc idl} package fits the object trace directly on the detector space; when
  the trace is too faint, it is interpolated from the adjacent orders based on
  offsets from a standard star trace. Optimal extraction is performed using a
  variant of the \citet{horne1986} algorithm.

\item Coaddition of spectra: The ESO pipeline coadds multiple ``nodding''
  exposures by aligning the object trace in the 2D rectified
  spectra. Coaddition of already rebinned spectra is not recommended, as it
  introduces a correlation between the error in adjacent pixels. Conversely,
  the {\sc idl} package does not attempt to add the 2D frames.  Instead, it
  optimally extracts a single 1D spectrum from all exposures in the same arm
  for a given object.

\item Ease of use: The ESO pipeline can be run automatically through the
  Reflex interface. The same is true for the {\sc idl} package, which is
  easily scriptable. One advantage of the latter is the possibility of
  obtaining both individual and co-added spectra from an arbitrarily large set
  of exposures in a single run, shortening the overall execution time.
\end{itemize}

\section{Post-processing}
\label{section_post}

In addition to the (approximately) flux calibrated spectra, we deliver to the
community two other higher level science data products: telluric-corrected
spectra and QSO continuum fits.

\subsection{Removal of telluric features}
\label{section_telluric}

\begin{figure}
\includegraphics[width=92mm,clip]{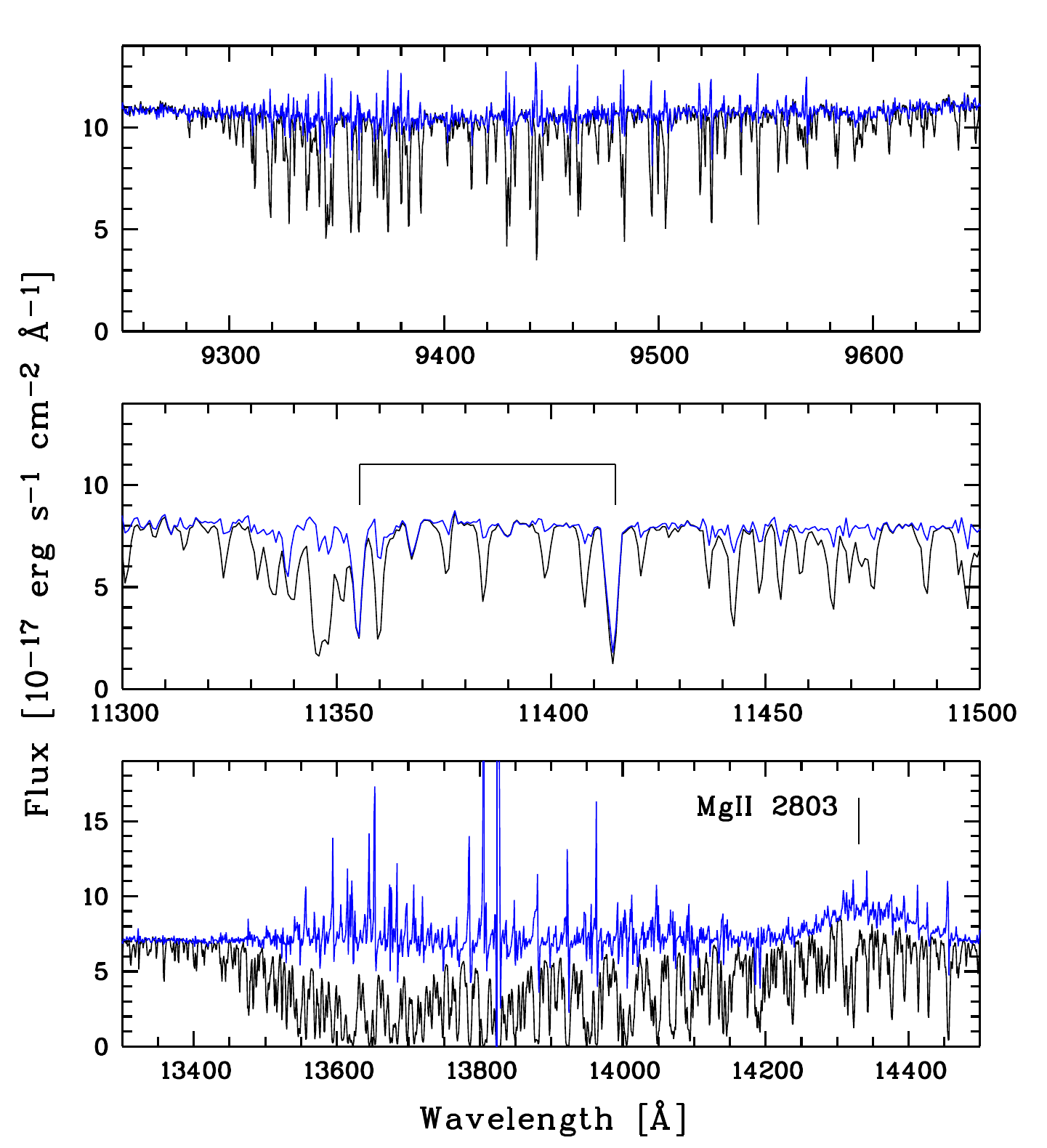}
\caption{Spectrum of QSO J0003$-$2603 at $z=4.125$ before (black) and after
  (blue) telluric corrections. Three spectral windows with strong telluric
  absorption are shown. In the middle panel the tickmarks above the spectrum
  indicate two \feii\ absorption lines, $\lambda2586$ and $\lambda2600$ \AA,
  associated with 
  the $z=3.390$ DLA.  In the bottom panel we note how the \mgii\ emission line
  stands out in the corrected spectrum.
\label{fig_telluric}}  
\end{figure}

Telluric absorption affects spectra in both the VIS and NIR arms. Correcting
these airmass-dependent spectral features using standard star spectra, even
taken relatively close in time to the science targets, can become highly
non-trivial owing to the rapidly changing NIR atmospheric transparency. Instead,
we opted to derive corrections using model transmission spectra based on the
ESO SKYCALC Cerro Paranal Advanced Sky Model~\citep{noll2012,jones2013},
version 1.3.5.  The SKYCALC models are a function of both airmass and
precipitable water vapor (PWV) and span a grid in these parameters providing a
spectral resolution of $R=100\,000$.  These corrections were applied to
individual-epoch spectra of all XQ-100 sources. Fig.~\ref{fig_telluric} shows
an example of the results.

Synthetic atmospheric transmission spectra based on the SKYCALC models were fit
separately to each VIS and NIR 1D spectrum as a way to remove the observed
telluric absorption features.  The sky model airmass and PWV parameters, as
well as a velocity offset and Gaussian FWHM smoothing kernel, were
interactively adjusted for each spectrum in order to minimize the residuals in
the model-subtracted spectrum over spectral regions observed to have moderate
amounts of absorption, e.g., $\sim$7150$-$7350\AA, $\sim$7620$-$7680\AA,
$\sim$8120$-$8350\AA, $\sim$8950$-$9250\AA, $\sim$9400$-$9600\AA,
$\sim$11000$-$11600\AA, and $\sim$14600$-$15000\AA.  Following this initial,
interactive parameter selection, an automated parameter selection was
performed that searched a grid of only airmass and PWV values in a narrow
grid, relative to the best-selected parameters from the interactive
search. Multiple sets of best-fit automated parameters were determined for
each spectrum by maximizing the S/N measured in the
model-subtracted VIS or NIR spectrum over each of the wavelength regions
listed above, separately, as well as an average S/N based on all VIS or NIR
regions, respectively.  The set of parameters used to create the final,
telluric-correction model was selected by eye from these multiple, best,
model-subtracted spectra.

Owing to the complex nature of correcting for the telluric absorption in this
way, which is affected by, for example, the degeneracy between fit parameters
and the variable atmospheric conditions during each observation, a single set
of parameters generally was not able to optimize the telluric absorption
correction at all wavelengths.  Similarly, a single quantitative measure of
``best'' was not attempted.  So while the final correction remains somewhat
subjective, this allowed an optimization of the correction over the wavelength
regions of greatest interest, e.g., near QSO emission lines, such as \civ\ or
\mgii\ (see bottom panel of Fig.~\ref{fig_telluric}) that will be important
for further analysis, and varies for objects at different redshifts.

The telluric correction models were fit to all available 1D spectra.  This
includes individual-epoch spectra for all objects, as well as spectra co-added
from multiple epochs.  We note, however, that these model-subtracted, co-added
spectra are of poorer quality than those from the individual epochs. This is a
result of the coadd of the multiple epochs being done to the
pre-telluric-corrected, 2D, unrectified images, which was done to avoid
rebinning multiple times.  However, this necessarily results in mixed
atmospheric features in the co-added spectrum.  Such features cannot be cleanly
fit by the atmospheric models.  In these cases, an argument can be made for
coadding the telluric-corrected, 1D spectra instead of the uncorrected 2D
frames, even if an additional rebinning is required; however, such
post-processing decisions and procedures are left to the user.

\subsection{Continuum fitting}
\label{section_continuum}

For each arm the manually placed continuum was
  determined by selecting points along the QSO continuum free of absorption
  (by eye) as knots for a cubic spline. The code used for the continuum
  fitting is available at \url{https://github.com/trystynb/ContFit}.

For all sightlines, the continuum placement was visually inspected and
adjusted such that the final fit resides within the variations of regions with
clean continuum. The accuracy of the fits is as good as or better than the S/N
of these clean continuum regions. As the continuum fits were created for
accurate DLA metal line abundances~\citep{berg2016}, the fits around
DLA metal lines have undergone multiple revisions compared to other regions of
the spectra.  The continuum placement in the Ly$\alpha$ forest is highly
subjective due to the lack of clean QSO continuum \citep[e.g.][]{kirkman2005},
and is particularly difficult to identify around the Ly$\alpha$ absorption of
DLAs. The continuum around a DLA Ly$\alpha$ absorption feature in the
XQ-100 sightlines requires further refinement on a case-by-case basis to match
the N(H\textsc{i}) fits of the Ly$\alpha$ wings, as implemented in
\citet{sanchez2015}.

In regions where the QSO continuum is absorbed, the spline knots were placed
at a constant (high) flux at: (i) The Lyman limit if one or more obvious Lyman
limits systems are clearly present, and (ii) telluric features (near observed
wavelengths 6900\AA{}, 7600\AA{}, 9450\AA{}, 11400\AA{}, and 14000\AA{}). In
some sightlines, there are strong absorption features on top of the Ly$\alpha$
emission line of the QSO, such that the continuum of the emission line is not
well constrained (particularly near the peak of the emission). In cases with
this strong absorption present, the continuum on the Ly$\alpha$ emission is
assumed to follow the interpolation from the cubic spline fit to the
surrounding continuum knots.

An example of the continuum presented above is shown in
Fig.~\ref{fig_spec}. We note, however, that we provide continua separately for
each arm-by-arm spectrum, not for the joint spectra.

\section{Description of science data products}
\label{section_products}

All the XQ-100 raw data, along with calibration files are available through
the ESO archive ({\tt http://archive.eso.org/eso/eso\_archive\_main.html}).
Advanced science data products (SDP) are publicly available in the form of ESO
Phase 3 material ({\tt http://archive.eso.org/wdb/wdb/adp/phase3\_main/form}).

The full XQ-100 target list is provided in Table~\ref{table_targets}.  We also
provide a summary file with basic properties (e.g., coordinates and redshifts),
spectroscopic properties (e.g.,  S/N at different rest frame wavelengths),
multi-wavelength photometric information, and other spectroscopic data
available for each XQ-100 QSO.  The detailed content of this summary file
is given in Table~\ref{table_parameters}. We format all the data file names in
the same fashion ({\tt JNNNNsNNNN}) and we provide this standardized name.

Two types of data are provided for each target: (1) individual UVB, VIS, and
NIR spectra, also with telluric correction and fitted QSO continuum; and (2)
a joint spectrum of the three arms together.

\subsection{Individual UVB, VIS, and NIR spectra}

There are four different main data files per QSO in the XQ-100 sample: one
with the reduced 1D spectrum in the UVB arm, one for the VIS arm, one for the
NIR reduced in ``stare'' mode, and one with the 1D NIR spectrum reduced in
nodding mode when available, i.e., when $z < 4$ (targets observed without the
K-band blocking filter).\footnote{The overall quality of the nodding reduction
  is worse than the normal reduction. Their unique advantage is that they
  extend up to the last NIR order; see Section~\ref{section_data_reduction}
  for details.}

Each spectrum file contains wavelength, flux, error on the flux,
sky-subtracted flux, and associated error (\S~\ref{section_telluric}).

When a target was observed more than once (because the observing
specifications were not met the first time; see
\S~\ref{sect_observations}), we produced individual spectra of each
execution of the OB.  Whenever possible, we also produced a co-added spectrum
putting together all executions. In the co-added spectra we discarded the
first exposures either when they were affected by the ADC issue, or when they
were interrupted (as their contribution was negligible due to the short
integration time). We define as “primary” spectra those with the best
achievable S/N. For targets observed more than once, these correspond to the
co-added spectra. 

A breakdown of the different spectra provided is shown in
Table~\ref{table_numberofspectra}.\\

\subsection{Joint spectra }

Joint spectra contain the three arms merged into a single spectrum.  Fluxes
from the VIS and NIR arms were rescaled to match the UVB flux level.  We first
computed the VIS scaling factor (using the UVB-VIS superposition); then, after
correcting the VIS, we computed the NIR scaling factor (using the VIS-NIR
superposition).  In both cases, the scaling factor was defined as the ratio
of the two median fluxes in the superposition region.  After rescaling, the
limit wavelength between UVB and VIS arms was set at \numprint{5600} \AA\
and at \numprint{10125} \AA\ between the VIS and NIR arms. The three arms
were finally pieced together to create a single spectrum.  For targets observed
without the K-band blocking filter, the last order of the NIR was taken from
the products of the nodding reduction, which are similarly rescaled, cut at
\numprint{22700} \AA, and pieced together.  The resulting spectrum was
finally cut in the blue end at \numprint{3000} \AA\ and in the red end at
\numprint{25000} \AA\ (for targets observed without the K-band blocking
filter) and at \numprint{18000} \AA\ (for other targets), to guarantee a
comparable wavelength span across the data set.  We note that the procedure
described above is the result of several choices that may not be appropriate
for all scientific analyses.

\subsection{Data format}

All the spectra we release are binary FITS files.  The naming convention is 

\begin{itemize}
\item for the individual arm-by-arm spectra: {\tt target\_{arm}\_{exec}.fits}
\item for the joint spectra: {\tt target.fits}
\end{itemize}

where {\tt target} is the target name in shortened J2000 coordinates
(JNNNN+NNNN or JNNNN$-$NNNN), {\tt arm} is the spectral arm, including the
optional nodding suffix for the NIR (uvb, vis, nir, or nir\_nod), and {\tt
  exec} is the optional execution suffix (\_1, \_2, \_3, or blank). The
individual arm-by-arm spectrum without the {\tt exec} suffix is to be regarded
as the primary spectrum for the given target in all cases.  The list of table
columns is

\begin{itemize}
\item for the individual arm-by-arm spectra: WAVE, FLUX, ERR\_FLUX, CONTINUUM,
  FLUX\_TELL\_CORR, ERR\_FLUX\_TELL\_CORR 
\item for the joint spectra: WAVE, FLUX, ERR\_FLUX 

\end{itemize}

The column description is as follows:

\begin{itemize}
\item WAVE: wavelength in the vacuum-heliocentric system (\AA);
\item FLUX: flux density (erg cm$^{-2}$ s$^{-1}$  Å$^{-1}$);
\item ERR\_FLUX: error of the flux density (erg cm$^{-2}$ s$^{-1}$  Å$^{-1}$);
\item CONTINUUM: fitted  continuum  (erg  cm$^{-2}$ s$^{-1}$  Å$^{-1}$);  
\item FLUX\_TELL\_CORR: same as flux, but with the telluric features removed
  (erg  cm$^{-2}$ s$^{-1}$  Å$^{-1}$);  
\item ERR\_FLUX\_TELL\_CORR: error of flux\_tc (erg cm$^{-2}$ s$^{-1}$  Å$^{-1}$)

\end{itemize}

\section{Summary}

We have presented XQ-100, a legacy survey of $100$ $z_{\rm em}\simeq3.5$--$4.5$
QSOs observed with VLT/XSHOOTER.  We have provided a basic description of the
sample, along with details of the observations, and details of the data
reduction process. We have also described the format and organization of the
publicly available data, which include spectra corrected for atmospheric
absorption and a continuum fit.

XQ-100 provides the first large uniform sample of high-redshift QSOs at
intermediate-resolution and with simultaneous rest-frame UV/optical
coverage. In terms of number of QSOs this volume represents a $30 \%$ increase
over the whole extant XSHOOTER sample.  The released spectra are of superb
quality, having median S/N $\sim 30$, $25$, and $40$ at resolutions of $\sim
30$--$50$ \kms, depending on wavelength.  We have indicated that these
properties enable a wide range of high-redshift research and soon  look
forward to seeing the results of this three-year effort in the form of
new discoveries and contributions to the field.

\section*{Acknowledgments}

We would like to warmly thank the ESO staff involved in the execution of this
Large Programme throughout all its phases.  SL has been supported by FONDECYT
grant number 1140838 and partially by PFB-06 CATA.  VD, IP, and SP acknowledge
support from the PRIN INAF 2012 ``The X-Shooter sample of 100 quasar spectra
at $z\sim 3.5$: Digging into cosmology and galaxy evolution with quasar
absorption lines''. SLE acknowledges the receipt of an NSERC Discovery
Grant. MH acknowledges support by ERC ADVANCED GRANT 320596 ``The Emergence of
Structure during the epoch of Reionization''. The Dark Cosmology Centre is
funded by the Danish National Research Foundation. MVe gratefully acknowledges
support from the Danish Council for Independent Research via grant no. DFF –
4002-00275. MV is supported by ERC-StG ``cosmoIGM''.  KDD is supported by an
NSF AAPF fellowship awarded under NSF grant AST-1302093. TSK acknowledges
funding support from the European Research Council Starting Grant ``Cosmology
with the IGM'' through grant GA-257670.

This research has made use
of the NASA/IPAC Extragalactic Database (NED), which is operated by the Jet
Propulsion Laboratory, California Institute of Technology, under contract with
the National Aeronautics and Space Administration.  Funding for the SDSS and
SDSS-II has been provided by the Alfred P. Sloan Foundation, the Participating
Institutions, the National Science Foundation, the U.S. Department of Energy,
the National Aeronautics and Space Administration, the Japanese
Monbukagakusho, the Max Planck Society, and the Higher Education Funding
Council for England. The SDSS Web Site is http://www.sdss.org/.  The SDSS is
managed by the Astrophysical Research Consortium for the Participating
Institutions. The Participating Institutions are the American Museum of
Natural History, Astrophysical Institute Potsdam, University of Basel,
University of Cambridge, Case Western Reserve University, University of
Chicago, Drexel University, Fermilab, the Institute for Advanced Study, the
Japan Participation Group, Johns Hopkins University, the Joint Institute for
Nuclear Astrophysics, the Kavli Institute for Particle Astrophysics and
Cosmology, the Korean Scientist Group, the Chinese Academy of Sciences
(LAMOST), Los Alamos National Laboratory, the Max-Planck-Institute for
Astronomy (MPIA), the Max-Planck-Institute for Astrophysics (MPA), New Mexico
State University, Ohio State University, University of Pittsburgh, University
of Portsmouth, Princeton University, the United States Naval Observatory, and
the University of Washington.

\appendix

\section{Tables}

\begin{onecolumn}
\small
\begin{longtable}{cc c c c c c c c }
  \caption{Summary of XQ-100 target properties}
\label{table_targets}\\
\hline\hline
XQ-100 name&NED Name &  RA &  DEC &  Redshift &  $R_{\rm APM}$ &  SNR$_{1700}$ &  SNR$_{3000}$ &  SNR$_{3600}$ \\ 
 (1) &(2) &(3) &(4) &(5) &(6) &(7)& (8) & (9) \\
\hline
\endfirsthead
\caption{continued.}\\
\hline\hline
XQ-100 name&NED Name &  RA &  DEC &  Redshift &  $R_{\rm APM}$  &  SNR$_{1700}$ &  SNR$_{3000}$ &  SNR$_{3600}$ \\ 
(1) &(2) &(3) &(4) &(5) &(6) &(7)& (8) & (9) \\
\hline
\endhead
\hline
\endfoot
J0003--2603 & HB89 0000--263              & 00 03 22.79 & --26 03 19.4 & 4.125 & 17.37 & 79 & 99 & -1 \\ 
J0006--6208 & BR J0006--6208              & 00 06 51.60 & --62 08 0.78 & 4.440 & 19.25 & 20 & 22 & -1 \\ 
J0030--5159 & BR J0030--5159              & 00 30 34.47 & --51 29 43.6 & 4.173 & 18.57 & 18 & 22 & -1 \\ 
J0034+1639 & PSS J0034+1639               & 00 34 54.71 & +16 39 18.2  & 4.292 & 18.03 & 28 & 30 & -1 \\ 
J0042--1020 & SDSS J004219.74--102009.4   & 00 42 19.73 & --10 20 12.2 & 3.863 & 18.23 & 52 & 48 & 58 \\ 
J0048--2442 & BRI J0048--2442             & 00 48 34.37 & --24 42 06.9 & 4.083 & 19.22 & 20 & 18 & -1 \\ 
J0056--2808 & HB89 0053--284              & 00 56 24.87 & --28 08 33.3 & 3.635 & 18.10 & 29 & 22 & 43 \\ 
J0057--2643 & HB89 0055--269              & 00 57 58.14 & --26 43 12.9 & 3.661 & 17.72 & 46 & 20 & 62 \\ 
J0100--2708 & PMN J0100--2708             & 01 00 12.47 & --27 08 52.1 & 3.546 & 18.87 & 30 & 6  & 30 \\ 
J0113--2803 & BRI J0113--2803             & 01 13 44.17 & --28 03 17.9 & 4.314 & 18.67 & 30 & 37 & -1 \\ 
J0117+1552 & PSS J0117+1552               & 01 17 31.05 & +15 52 14.2  & 4.243 & 17.22 & 40 & 62 & -1 \\ 
J0121+0347 & PSS J0121+0347               & 01 21 26.21 & +03 47 04.7  & 4.125 & 18.33 & 31 & 30 & -1 \\ 
J0124+0044 & SDSS J0124+0044              & 01 24 03.97 & +00 44 31.4  & 3.837 & 17.75 & 34 & 41 & 48 \\ 
J0132+1341 & PSS J0132+1341               & 01 32 09.98 & +13 41 35.9  & 4.152 & 18.53 & 32 & 30 & -1 \\ 
J0134+0400 & PSS J0134+0400               & 01 33 40.47 & +04 00 58.5  & 4.185 & 18.32 & 48 & 52 & -1 \\ 
J0137--4224 & BRI J0137--4224             & 01 37 24.36 & --42 24 14.9 & 3.971 & 18.77 & 17 & 18 & 17 \\ 
J0153--0011 & SDSS J015339.60--001104.8   & 01 53 39.73 & --00 11 06.1 & 4.195 & 18.87 & 15 & 18 & -1 \\ 
J0211+1107 & PSS J0211+1107               & 02 11 20.10 & +11 07 14.5  & 3.973 & 18.20 & 22 & 26 & 25 \\ 
J0214--0518 & PMN J0214--0518             & 02 14 29.41 & --05 17 45.4 & 3.977 & 18.42 & 31 & 28 & 24 \\ 
J0234--1806 & BR J0234--1806              & 02 34 55.03 & --18 06 11.3 & 4.305 & 18.79 & 28 & 30 & -1 \\ 
J0244--0134 & BRI 0241--0146              & 02 44 01.83 & --01 34 06.3 & 4.055 & 18.18 & 39 & 44 & -1 \\ 
J0247--0555 & BR 0245--0608               & 02 47 56.70 & --05 56 00.0 & 4.234 & 18.65 & 22 & 29 & -1 \\ 
J0248+1802 & PSS J0248+1802               & 02 48 54.37 & +18 02 47.0  & 4.439 & 17.71 & 26 & 40 & -1 \\ 
J0255+0048 & SDSS J025518.57+004847.4     & 02 55 18.70 & +00 48 46.5  & 4.003 & 18.31 & 30 & 32 & 22 \\ 
J0307--4945 & BR J0307--4945              & 03 07 22.57 & --49 45 45.6 & 4.716 & 18.76 & 37 & 82 & -1 \\ 
J0311--1722 & BR J0311--1722              & 03 11 15.38 & --17 22 48.4 & 4.034 & 17.73 & 39 & 37 & -1 \\ 
J0401--1711 & BR J0401--1711              & 04 03 56.82 & --17 03 22.0 & 4.227 & 18.69 & 21 & 28 & -1 \\ 
J0415--4357 & BR J0415--4357              & 04 15 15.18 & --43 57 50.7 & 4.073 & 18.81 & 16 & 28 & -1 \\ 
J0424--2209 & BR J0424--2209              & 04 26 10.47 & --22 02 17.5 & 4.329 & -1    & 26 & 33 & -1 \\ 
J0523--3345 & BR J0523--3345              & 05 25 05.95 & --33 43 4.44 & 4.385 & 18.37 & 39 & 65 & -1 \\ 
J0529--3526 & BR J0529--3526              & 05 29 15.98 & --35 26 01.2 & 4.418 & 18.94 & 22 & 25 & -1 \\ 
J0529--3552 & BR J0529--3552              & 05 29 20.94 & --35 52 31.8 & 4.172 & 18.29 & 13 & 14 & -1 \\ 
J0714--6455 & BR J0714--6455              & 07 14 30.92 & --64 55 10.3 & 4.465 & 18.35 & 29 & 48 & -1 \\ 
J0747+2739 & SDSS J074711.15+273903.3     & 07 47 11.17 & +27 39 00.8  & 4.133 & 17.24 & 27 & 34 & -1 \\ 
J0755+1345 & SDSS J075552.41+134551.1     & 07 55 52.43 & +13 45 49.6  & 3.663 & 18.75 & 29 & 9  & 32 \\ 
J0800+1920 & SDSS J080050.27+192058.9     & 08 00 50.26 & +19 20 56.3  & 3.948 & 18.27 & 29 & 28 & 33 \\ 
J0818+0958 & SDSS J081855.78+095848.0     & 08 18 55.75 & +09 58 44.9  & 3.656 & 17.69 & 38 & 1  & 44 \\ 
J0833+0959 & SDSS J083322.50+095941.2     & 08 33 22.50 & +09 59 38.6  & 3.716 & 18.52 & 33 & 13 & 37 \\ 
J0835+0650 & SDSS J083510.92+065052.8     & 08 35 10.91 & +06 50 51.0  & 4.007 & 17.95 & 33 & 34 & 20 \\ 
J0839+0318 & SDSS J083941.45+031817.0     & 08 39 41.58 & +03 18 18.2  & 4.230 & 17.85 & 12 & 19 & -1 \\ 
J0920+0725 & SDSS J092041.76+072544.0     & 09 20 41.72 & +07 25 41.2  & 3.646 & 18.53 & 40 & 3  & 37 \\ 
J0935+0022 & SDSS J093556.91+002255.6     & 09 35 56.87 & +00 22 52.8  & 3.747 & 17.78 & 27 & 15 & 25 \\ 
J0937+0828 & SDSS J093714.48+082858.6     & 09 37 14.51 & +08 28 56.2  & 3.704 & 18.15 & 23 & 4  & 37 \\ 
J0955--0130 & BRI 0952--0115              & 09 55 00.01 & --01 30 08.4 & 4.418 & 18.66 & 35 & 37 & -1 \\ 
J0959+1312 & SDSS J095937.11+131215.4     & 09 59 37.23 & +13 12 17.6  & 4.092 & 16.87 & 54 & 76 & -1 \\ 
J1013+0650 & J101347+065015               & 10 13 47.48 & +06 50 16.6  & 3.809 & 18.38 & 30 & 22 & 38 \\ 
J1018+0548 & J101818+054822               & 10 18 18.57 & +05 48 20.7  & 3.515 & 18.14 & 30 & 14 & 38 \\ 
J1020+0922 & J102040+092254               & 10 20 40.74 & +09 22 53.1  & 3.640 & 18.02 & 22 & 4  & 19 \\ 
J1024+1819 & SDSSJ1024+1819               & 10 24 56.78 & +18 19 07.1  & 3.524 & 17.89 & 26 & 2  & 27 \\ 
J1032+0927 & J103221+092748               & 10 32 21.26 & +09 27 47.5  & 3.985 & 17.94 & 27 & 22 & 17 \\ 
J1034+1102 & J103446+110214               & 10 34 46.55 & +11 02 12.0  & 4.269 & 18.18 & 33 & 35 & -1 \\ 
J1036--0343 & BR 1033--0327               & 10 36 23.63 & --03 43 21.0 & 4.531 & 19.18 & 19 & 47 & -1 \\ 
J1037+2135 & SDSSJ1037+2135               & 10 37 30.43 & +21 35 29.8  & 3.626 & 17.69 & 52 & 11 & 62 \\ 
J1037+0704 & J103732+070426               & 10 37 32.31 & +07 04 23.7  & 4.127 & 18.33 & 49 & 46 & -1 \\ 
J1042+1957 & SDSSJ1042+1957               & 10 42 34.02 & +19 57 16.3  & 3.630 & 18.11 & 36 & 12 & 37 \\ 
J1053+0103 & SDSS J105340.75+010335.6     & 10 53 40.82 & +01 03 33.5  & 3.663 & 19.11 & 36 & 8  & 37 \\ 
J1054+0215 & J105434+021551               & 10 54 34.33 & +02 15 51.3  & 3.971 & 18.03 & 14 & 13 & 12 \\ 
J1057+1910 & J105705+191042               & 10 57 05.53 & +19 10 43.7  & 4.128 & 17.87 & 19 & 21 & -1 \\ 
J1058+1245 & J105858+124554               & 10 58 58.51 & +12 45 53.8  & 4.341 & 17.64 & 26 & 35 & -1 \\ 
J1103+1004 & J110352+100403               & 11 03 52.72 & +10 04 0.48  & 3.607 & 18.61 & 44 & 14 & 64 \\ 
J1108+1209 & J110855+120953               & 11 08 55.56 & +12 09 51.7  & 3.679 & 18.40 & 40 & 10 & 62 \\ 
J1110+0244 & J111008+024458               & 11 10 08.81 & +02 44 57.3  & 4.146 & 17.59 & 30 & 36 & -1 \\ 
J1111--0804 & BRI 1108--0747              & 11 11 13.89 & --08 04 03.9 & 3.922 & 18.82 & 43 & 38 & 55 \\ 
J1117+1311 & J111701+131115               & 11 17 01.97 & +13 11 13.0  & 3.622 & 18.28 & 39 & 9  & 47 \\ 
J1126--0126 & J112617--012632             & 11 26 17.54 & --01 26 34.2 & 3.635 & 18.70 & 22 & 4  & 24 \\ 
J1126--0124 & J112634--012436             & 11 26 34.42 & --01 24 38.0 & 3.765 & 18.53 & 27 & 15 & 21 \\ 
J1135+0842 & J113536+084218               & 11 35 36.55 & +08 42 17.3  & 3.834 & 18.26 & 55 & 49 & 53 \\ 
J1201+1206 & HB89 1159+123                & 12 01 48.05 & +12 06 28.2  & 3.522 & 17.32 & 52 & 4  & 83 \\ 
J1202--0054 & SDSSJ1202--0054             & 12 02 10.06 & --00 54 27.9 & 3.592 & 18.49 & 23 & 3  & 23 \\ 
J1248+1304 & J124837+130440               & 12 48 37.39 & +13 04 39.2  & 3.721 & 18.14 & 39 & 22 & 53 \\ 
J1249--0159 & J124957--015928             & 12 49 57.40 & --01 59 29.8 & 3.629 & 17.47 & 37 & 18 & 46 \\ 
J1304+0239 & J130452+023924               & 13 04 52.60 & +02 39 21.8  & 3.648 & 18.55 & 47 & 13 & 54 \\ 
J1312+0841 & J131242+084105               & 13 12 42.94 & +08 41 02.8  & 3.731 & 18.41 & 33 & 30 & 46 \\ 
J1320--0523 & J1320299--052335            & 13 20 30.12 & --05 23 36.3 & 3.717 & 17.81 & 41 & 18 & 65 \\ 
J1323+1405 & J132346+140517               & 13 23 46.21 & +14 05 16.4  & 4.054 & 18.60 & 23 & 21 & -1 \\ 
J1330--2522 & BR J1330--2522              & 13 30 52.17 & --25 22 18.1 & 3.949 & 18.46 & 39 & 45 & 44 \\ 
J1331+1015 & SDSS J133150.69+101529.4     & 13 31 50.77 & +10 15 27.5  & 3.852 & 18.76 & 33 & 33 & 40 \\ 
J1332+0052 & J133254+005250               & 13 32 54.60 & +00 52 48.3  & 3.508 & 18.43 & 41 & 17 & 57 \\ 
J1336+0243 & J133653+024338               & 13 36 53.43 & +02 43 35.5  & 3.801 & 18.62 & 33 & 20 & 36 \\ 
J1352+1303 & J135247+130311               & 13 52 48.09 & +13 03 09.8  & 3.706 & 18.35 & 14 & 3  & 15 \\ 
J1401+0244 & J1401+0244                   & 14 01 46.52 & +02 44 37.7  & 4.408 & 18.41 & 39 & 47 & -1 \\ 
J1416+1811 & SDSSJ1416+1811               & 14 16 08.32 & +18 11 46.1  & 3.593 & 18.19 & 24 & 6  & 23 \\ 
J1421--0643 & PKS B1418--064              & 14 21 07.93 & --06 43 57.6 & 3.688 & 19.03 & 40 & 17 & 45 \\ 
J1442+0920 & J144250+092001               & 14 42 50.12 & +09 19 58.9  & 3.532 & 17.21 & 42 & 7  & 46 \\ 
J1445+0958 & SDSSJ1445+0958               & 14 45 16.62 & +09 58 34.9  & 3.562 & 17.64 & 40 & 4  & 43 \\ 
J1503+0419 & J150328+041949               & 15 03 29.01 & +04 19 47.3  & 3.692 & 18.01 & 41 & 18 & 45 \\ 
J1517+0511 & SDSSJ1517+0511               & 15 17 56.20 & +05 11 00.7  & 3.555 & 18.31 & 41 & 5  & 38 \\ 
J1524+2123 & SDSSJ1524+2123               & 15 24 36.17 & +21 23 07.0  & 3.600 & 17.25 & 27 & 8  & 42 \\ 
J1542+0955 & J154237+095558               & 15 42 37.62 & +09 56 01.2  & 3.986 & 18.18 & 31 & 24 & 16 \\ 
J1552+1005 & J155255+100538               & 15 52 55.22 & +10 05 37.0  & 3.722 & 18.63 & 35 & 11 & 49 \\ 
J1621--0042 & J1621--0042                 & 16 21 17.04 & --00 42 52.9 & 3.711 & 17.67 & 34 & 27 & 77 \\ 
J1633+1411 & J163319+141142               & 16 33 19.69 & +14 11 39.7  & 4.365 & 18.72 & 31 & 45 & -1 \\ 
J1658--0739 & J1658--0739                 & 16 58 44.20 & --07 39 16.4 & 3.750 & -1    & 37 & 45 & 78 \\ 
J1723+2243 & PSS J1723+2243               & 17 23 23.13 & +22 43 54.7  & 4.531 & 18.71 & 16 & 99 & -1 \\ 
J2215--1611 & BR 2212--1626               & 22 15 27.26 & --16 11 34.3 & 3.995 & -1    & 40 & 54 & 45 \\ 
J2216--6714 & BR 2213--6729               & 22 16 51.98 & --67 14 41.2 & 4.479 & 18.57 & 21 & 40 & -1 \\ 
J2239--0552 & J2239536--055219            & 22 39 53.62 & --05 52 21.3 & 4.557 & 18.30 & 10 & 26 & -1 \\ 
J2251--1227 & BR 2248--1242               & 22 51 18.19 & --12 27 05.1 & 4.157 & 18.55 & 34 & 58 & -1 \\ 
J2344+0342 & PSS J2344+0342               & 23 44 03.05 & +03 42 24.3  & 4.248 & 18.16 & 32 & 33 & -1 \\ 
J2349--3712 & BR J2349--3712              & 23 49 13.56 & --37 12 59.8 & 4.219 & 19.19 & 21 & 29 & -1 \\ 
\hline
\end{longtable}
\end{onecolumn}
Columns: 


\begin{inparaenum}[(1)]

\item Target name used throughout  this paper and also given to the files
  associated with each object; 
\item target's NED name, also found in the SDP headers; 
\item right ascension in sexagesimal degrees (J2000); 
\item declination in sexagesimal degrees (J2000);
\item QSO redshift estimated using the result of a principal component
analysis \citep{paris2012}; 
\item APM $R$-magnitude. When the magnitude was not
found, the value was set to $-1$; 
\item , 
\item  and 
\item  average pixel SNR in the co-added spectrum near rest-frame
wavelengths $1\,700$,
$3\,000$ and $3\,600$ \AA,  respectively; set to $-1$ if wavelength was not
covered, i.e.,  for spectra taken with the K-band blocking filter ($z_{\rm
  em}>4$ 
sources). 

\end{inparaenum}

\newpage

\begin{onecolumn}
\begin{longtable}{c l c l} 
\caption{Parameters associated with each XQ-100 object in the public
  repository}
\label{table_parameters}\\ 
\hline\hline
Column &Name &Format &Description\\
\hline
\endfirsthead
\caption{continued.}\\
\hline\hline
Column &Name &Format &Description\\
\hline
\endhead
\hline
\endfoot
1 & OBJECT & STRING & target designation\\
2 & RA\_J2000 & DOUBLE & target right ascension (deg, J2000.0)\\
3 & DEC\_J2000 & DOUBLE & target declination (deg, J2000.0)\\
4 & Z\_QSO & FLOAT & quasar emission redshift (PCA)\\
5 & N\_OBS & SHORT & number of observing epochs\\
6 & MJD\_OBS & FLOAT & start of observations (d)\\
7 & MJD\_OBS\_1 & FLOAT & start of observations (1st exec. only) (d)\\
8 & MJD\_OBS\_2 & FLOAT & start of observations (2nd exec. only) (d)\\
9 & MJD\_OBS\_3 & FLOAT & start of observations (3rd exec. only) (d)\\
10 & MJD\_END & FLOAT & end of observations (d)\\
11 & MJD\_END\_1 & FLOAT & end of observations (1st exec. only) (d)\\
12 & MJD\_END\_2 & FLOAT & end of observations (2nd exec. only) (d)\\
13 & MJD\_END\_2 & FLOAT & end of observations (3rd exec. only) (d)\\
14 & SEEING\_MIN & FLOAT & min. seeing from ESO.TEL.IA.FWHM keyw.\\
15 & SEEING\_MIN\_1 & FLOAT & min. seeing from ESO.TEL.IA.FWHM keyw. (1st exec. only)\\
16 & SEEING\_MIN\_2 & FLOAT & min. seeing from ESO.TEL.IA.FWHM keyw. (2nd exec. only)\\
17 & SEEING\_MIN\_3 & FLOAT & min. seeing from ESO.TEL.IA.FWHM keyw. (3rd exec. only)\\
18 & SEEING\_MAX & FLOAT & max. seeing measured at the start or at the end of integrations\\
19 & SEEING\_MAX\_1 & FLOAT & max. seeing from ESO.TEL.IA.FWHM keyw. (1st exec. only)\\
20 & SEEING\_MAX\_2 & FLOAT & max. seeing from ESO.TEL.IA.FWHM keyw. (2nd exec. only)\\
21 & SEEING\_MAX\_3 & FLOAT & max. seeing from ESO.TEL.IA.FWHM keyw. (3rd exec. only)\\
22 & SNR\_170 & FLOAT & S/N in a $\pm $1 nm window at 170 nm (rest-frame)\\
23 & SNR\_170\_1 & FLOAT & S/N in a $\pm $1 nm window at 170 nm (1st exec. only) (rest-frame)\\
24 & SNR\_170\_2 & FLOAT & S/N in a $\pm $1 nm window at 170 nm (2nd exec. only) (rest-frame)\\
25 & SNR\_170\_3 & FLOAT & S/N in a $\pm $1 nm window at 170 nm (3rd exec. only) (rest-frame)\\
26 & SNR\_300 & FLOAT & S/N in a $\pm $1 nm window at 300 nm (rest-frame)\\
27 & SNR\_300\_1 & FLOAT & S/N in a $\pm $1 nm window at 300 nm (1st exec. only) (rest-frame)\\
28 & SNR\_300\_2 & FLOAT & S/N in a $\pm $1 nm window at 300 nm (2nd exec. only) (rest-frame)\\
29 & SNR\_300\_3 & FLOAT & S/N in a $\pm $1 nm window at 300 nm (3rd exec. only) (rest-frame)\\
30 & SNR\_360 & FLOAT & S/N in a $\pm $1 nm window at 360 nm (rest-frame)\\
31 & SNR\_360\_1 & FLOAT & S/N in a $\pm $1 nm window at 360 nm (1st exec. only) (rest-frame)\\
32 & SNR\_360\_2 & FLOAT & S/N in a $\pm $1 nm window at 360 nm (2nd exec. only) (rest-frame)\\
33 & SNR\_360\_3 & FLOAT & S/N in a $\pm $1 nm window at 360 nm (3rd exec. only) (rest-frame)\\
34 & RED\_QUAL & SHORT & reduction quality parameter (see above)\\
35 & RED\_QUAL\_1 & SHORT & reduction quality parameter (1st exec. only)\\
36 & RED\_QUAL\_2 & SHORT & reduction quality parameter (2nd exec. only)\\
37 & RED\_QUAL\_3 & SHORT & reduction quality parameter (3rd exec. only)\\
38 & HR\_FLAG & SHORT & high-resolution spectrum flag\\
39 & JOHNSON\_MAG\_B & FLOAT & B magnitudes in Johnson system\\
40 & JOHNSON\_MAG\_V & FLOAT & V magnitudes in Johnson system\\
41 & JOHNSON\_MAG\_R & FLOAT & R magnitudes in Johnson system\\
42 & SDSS\_PSF\_MAG\_u & DOUBLE & SDSS PSF u magnitudes\\
43 & SDSS\_ERR\_PSF\_MAG\_u & DOUBLE & Error on SDSS PSF u magnitudes\\
44 & SDSS\_PSF\_MAG\_g & DOUBLE & SDSS PSF g magnitudes\\
45 & SDSS\_ERR\_PSF\_MAG\_g & DOUBLE & Error on SDSS PSF g magnitudes\\
46 & SDSS\_PSF\_MAG\_r & DOUBLE & SDSS PSF r magnitudes\\
47 & SDSS\_ERR\_PSF\_MAG\_r & DOUBLE & Error on SDSS PSF r magnitudes\\
48 & SDSS\_PSF\_MAG\_i & DOUBLE & SDSS PSF i magnitudes\\
49 & SDSS\_ERR\_PSF\_MAG\_i & DOUBLE & Error on SDSS PSF i magnitudes\\
50 & SDSS\_PSF\_MAG\_z & DOUBLE & SDSS PSF z magnitudes\\
51 & SDSS\_ERR\_PSF\_MAG\_z & DOUBLE & Error on SDSS PSF z magnitudes\\
52 & DR7Q\_MATCH & SHORT & match in DR7Q spectroscopy\\
53 & DR7Q\_LATE & INT32 & DR7Q plate number\\
54 & DR7Q\_MJD & INT32 & DR7Q spectroscopic MJD (d)\\
55 & DR7Q\_FIBER & INT32 & DR7Q fiber number (d)\\
56 & DR12Q\_MATCH & SHORT & match in DR12Q spectroscopy\\
57 & DR12Q\_N & INT32 & number of spectroscopic observations in DR12Q\\
58 & DR12Q\_PLATE\_1 & INT32 & DR12Q plate number (1st observation)\\
59 & DR12Q\_MJD\_1 & INT32 & DR12Q spectroscopic MJD (1st observation) (d)\\
60 & DR12Q\_FIBER\_1 & INT32 & DR12Q fiber number (1st observation) (d)\\
61 & DR12Q\_PLATE\_2 & INT32 & DR12Q plate number (2nd observation)\\
62 & DR12Q\_MJD\_2 & INT32 & DR12Q spectroscopic MJD (2nd observation) (d)\\
63 & DR12Q\_FIBER\_2 & INT32 & DR12Q fiber number (2nd observation) (d)\\
64 & FIRST\_MATCH & INT32 & match in FIRST\\
65 & FIRST\_FLUX & DOUBLE & FIRST flux at 20 cm (mJy)\\
66 & FIRST\_SNR & DOUBLE & S/N of FIRST detection\\
67 & TMASS\_MATCH & SHORT & matched in 2MASS\\
68 & TMASS\_MAG\_J & DOUBLE & 2MASS J magnitudes\\
69 & TMASS\_ERR\_MAG\_J & DOUBLE & error on 2MASS J magnitudes\\
70 & TMASS\_SNR\_J & DOUBLE & S/N of 2MASS detection in J bands\\
71 & TMASS\_MAG\_H & DOUBLE & 2MASS H magnitudes\\
72 & TMASS\_ERR\_MAG\_H & DOUBLE & error on 2MASS H magnitudes\\
73 & TMASS\_SNR\_H & DOUBLE & S/N of 2MASS detection in H bands\\
74 & TMASS\_MAG\_K & DOUBLE & 2MASS K magnitudes\\
75 & TMASS\_ERR\_MAG\_K & DOUBLE & error on 2MASS K magnitudes\\
76 & TMASS\_SNR\_K & DOUBLE & S/N of 2MASS detection in K bands\\
77 & TMASS\_RD\_FLAG & STRING & 2MASS rd flag\\
78 & WISE\_MATCH & SHORT & match in WISE\\
79 & WISE\_MAG\_w1 & DOUBLE & WISE w1 magnitudes\\
80 & WISE\_ERR\_MAG\_w1 & DOUBLE & error on WISE w1 magnitudes\\
81 & WISE\_SNR\_w1 & DOUBLE & S/N of WISE detection in w1 bands\\
82 & WISE\_RCHI2\_w1 & DOUBLE & WISE reduced chi-squared in w1 bands\\
83 & WISE\_MAG\_w2 & DOUBLE & WISE w2 magnitudes\\
84 & WISE\_ERR\_MAG\_w2 & DOUBLE & error on WISE w2 magnitudes\\
85 & WISE\_SNR\_w2 & DOUBLE & S/N of WISE detection in w2 bands\\
86 & WISE\_RCHI2\_w2 & DOUBLE & WISE reduced chi-squared in w2 bands\\
87 & WISE\_MAG\_w3 & DOUBLE & WISE w3 magnitudes\\
88 & WISE\_ERR\_MAG\_w3 & DOUBLE & error on WISE w3 magnitudes\\
89 & WISE\_SNR\_w3 & DOUBLE & S/N of WISE detection in w3 bands\\
90 & WISE\_RCHI2\_w3 & DOUBLE & WISE reduced chi-squared in w3 bands\\
91 & WISE\_MAG\_w4 & DOUBLE & WISE w4 magnitudes\\
92 & WISE\_ERR\_MAG\_w4 & DOUBLE & error on WISE w4 magnitudes\\
93 & WISE\_SNR\_w4 & DOUBLE & S/N of WISE detection in w4 bands\\
94 & WISE\_RCHI2\_w4 & DOUBLE & WISE reduced chi-squared in w4 bands\\
95 & WISE\_CC\_FLAG & STRING & WISE confusion and contamination flag\\
96 & WISE\_PH\_QUAL & STRING & WISE photometric quality flag \\ 
\hline
\end{longtable}
\end{onecolumn}

Notes on the catalog columns:\\

\noindent
1. Object name as designated in the ESO archive.

\noindent
2-3. The J2000 coordinates (Right Ascension and Declination) in sexagesimal degrees.  

\noindent
4. QSO redshift. The redshift was estimated using the result of a principal component analysis \citep{paris2012}.

\noindent
5. Number of XSHOOTER observations. Most QSOs were observed only
once. Thirteen  QSOs were observed more than once because
of interrupted OBs or ADC issues.

\noindent
6-9. Modified Julian Day (MJD) at the beginning of XSHOOTER observation. The values for the different executions of the same observing block are also listed separately (when applicable).

\noindent
10-13. Modified Julian Day (MJD) at the end of XSHOOTER observation. The values for the different executions of the same observing block are also listed separately (when applicable).

\noindent
14-17. Minimum seeing of XSHOOTER observation, taken from the ESO.TEL.IA.FWHM keyword, expressed in arcsec. The values for the different executions of the same observing block are also listed separately (when applicable).

\noindent
18-21. Minimum seeing of XSHOOTER observation, taken from the ESO.TEL.IA.FWHM keyword, expressed in arcsec. The values for the different executions of the same observing block are also listed separately (when applicable).

\noindent
22-25. Average S/N near \numprint{1700} \AA\ (rest frame) computed in the window \numprint{1690}-\numprint{1710}\AA. The values for the different executions of the same observing block are also listed separately (when applicable) .

\noindent
26-29. Average S/N near \numprint{3000} \AA\ (rest frame) computed in the window \numprint{2990}-\numprint{3010}\AA. The values for the different executions of the same observing block are also listed separately (when applicable).

\noindent
30-33. Average S/N near \numprint{3600} \AA\ (rest frame) computed in the window \numprint{3590}-\numprint{3610}\AA. The values for the different executions of the same observing block are also listed separately (when applicable).

\noindent
34-37. Calibration flags for each XSHOOTER observation. The value of the
resulting calibration flag is the sum of the five following flags.  A value of
0 means no problem to report, 1 means that the VIS spectrum was calibrated
using a different standard star, 2 means that there are residual
spikes in the UVB spectrum, 4 is set when apparent order-to-order fluctuations
in the VIS arm, 8 when the exposure was interrupted, and a value of 16 is set
when the exposure was taken with faulty ADCs. The values for the different executions of the same observing block (when applicable) are also listed separately.

\noindent
38. High-resolution spectroscopy (Keck/HIRES or VLT/UVES) exist for some of the XQ-100 QSOs. The {\tt HR\_FLAG} is set to 1 if a high-resolution
spectrum exists, otherwise 0.

\noindent
39-41.. Magnitudes in the $B$, $V$ and $R$ Johnson filters. These values were
retrieved from the CDS ({\it Centre de Données astronomiques de Strasbourg}). When the magnitude in one of the
filters could not be found, the value was set to $-1$.

\noindent
42-51. SDSS-DR12 point-spread function magnitudes (Cols.\#42, 44, 46, 49, 50) and their associated errors (Cols.\#43, 45, 47, 48, 51) in the $u$, $g$, $r$, $i$ and $z$ filters \citep{DR12}. Objects outside of the SDSS
footprint have associated magnitudes and errors set to $-1$.

\noindent
52. If a QSO was observed as part of SDSS-I/II \citep{york2000,schneider2010},
the {\tt DR7Q\_MATCHED} flag is set to 1, otherwise 0.

\noindent
53-55. When a SDSS-I/II spectrum is available, the SDSS plate number (Col.\#53), spectroscopic MJD (Col.\#54) and fiber number (Col.\#55).

\noindent
56. If a QSO was observed as part of SDSS-III \citep{eisenstein2011}, the {\tt
  DR12Q\_MATCHED} flag is set to 1, otherwise 0.

\noindent
57. Number of SDSS-III spectra available.

\noindent
58-63. When SDSS-III spectra are available, the plate numbers (Cols.\#58, 61), spectroscopic MJDs (Cols.\#59, 62) and fiber numbers (Col.\#60, 63). The values of the first and second observation are listed separately (when applicable).

\noindent
64. If there is a source in the FIRST radio catalog \citep[version March 2014;
][]{becker1995} within 5\arcsec\ of the QSO position, the {\tt FIRST\_MATCHED}
flag is set to 1, otherwise 0. If the QSO lies outside of the FIRST footprint,
it is set to $-1$.

\noindent
65. FIRST peak flux density at 20 cm, expressed in mJy.

\noindent
66. S/N of the FIRST source whose flux is given in Col.\#65.

\noindent
67. If there is a source from the Two Micron All Sky Survey All-Sky Data
Release Point Source Catalog \citep[2MASS; ][]{cutri2003} within 5\arcsec\ of
the QSO position, the {\tt TMASS\_MATCHED} is set to 1, otherwise 0.

\noindent
68-76. $J$, $H$, and $K$ magnitudes (Cols.\#68, 71, 74), with their associated error (Cols.\#69, 72, 75) and S/N (Cols.\#70, 73, 76). We note that 2MASS magnitudes are Vega-based.

\noindent
77. 2MASS {\tt rd\_flag} gives the meaning of the peculiar values of the
magnitudes and errors\footnote{see
  http://www.ipac.caltech.edu/2mass/releases/allsky/doc/explsup.html}.

\noindent
78. If a source from the Wide-field Infrared Survey Explorer AllWISE Data
Release Point Source Catalog \citep[WISE; ][]{wright2010} lies within
5\arcsec\ of a XQ-100 QSO, the {\tt WISE\_MATCHED} is set to 1, otherwise 0.

\noindent
79-94. WISE $w1$, $w2$, $w3$, and $w4$ magnitudes (Cols\#79, 83, 87, 91), with their associated errors (Cols\#80, 84, 88, 92), S/N (Cols\#81, 85, 89, 93) and $\chi ^2$ (Cols\#82, 86, 90, 94).

\noindent 
95. WISE contamination and confusion flag.

\noindent
96. WISE photometric quality flag.

\newpage

\begin{table*}
 \centering
 \begin{minipage}{160mm}
  \caption{Number of reduced spectra\label{table_numberofspectra}}
  \begin{tabular}{lcccccc}

  \hline

      & UVB & VIS  & NIR & NIR (nodded) & Merged & Total\\ 

 \hline

 Primary          & 100 & 100 & 100 & 47 & 100 & \\ 
 First execution  &  8  &  8  &  8  &  5 & --  & \\ 
 Second execution &  8  &  8  &  8  &  5 & --  &\\  
 Third execution  &  2  &  2  &  2  &  1 & --  &\\  
 Total            &  118& 118 & 118 &  58& 100  & 512\\  

\hline

\end{tabular}
\end{minipage}
\end{table*}

\newpage

\section{Spectra}

\begin{figure*}
\includegraphics[width=170mm,clip]{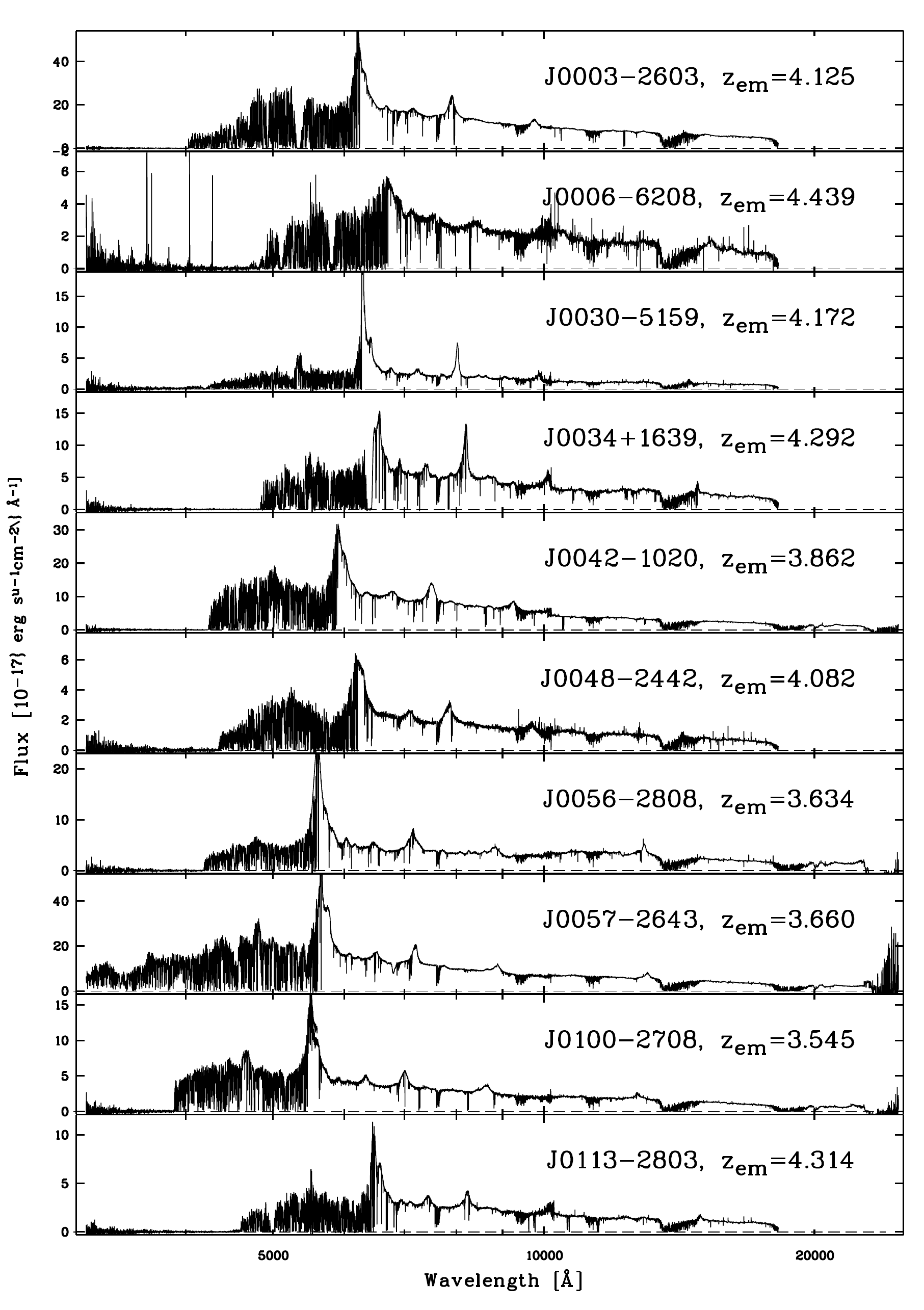}
\caption{XQ-100 spectra. Names follow the XQ-100 convention
  (\S~\ref{section_products}); see Table~\ref{table_targets} for a
  correspondence with literature names.  Emission redshifts were estimated
  using the result of a principal component analysis \citep{paris2012}.  The
  flux has been smoothed with a five-pixel median filter for displaying
  purposes.
\label{fig_all}}
\end{figure*}

\begin{figure*}
\includegraphics[width=170mm,clip]{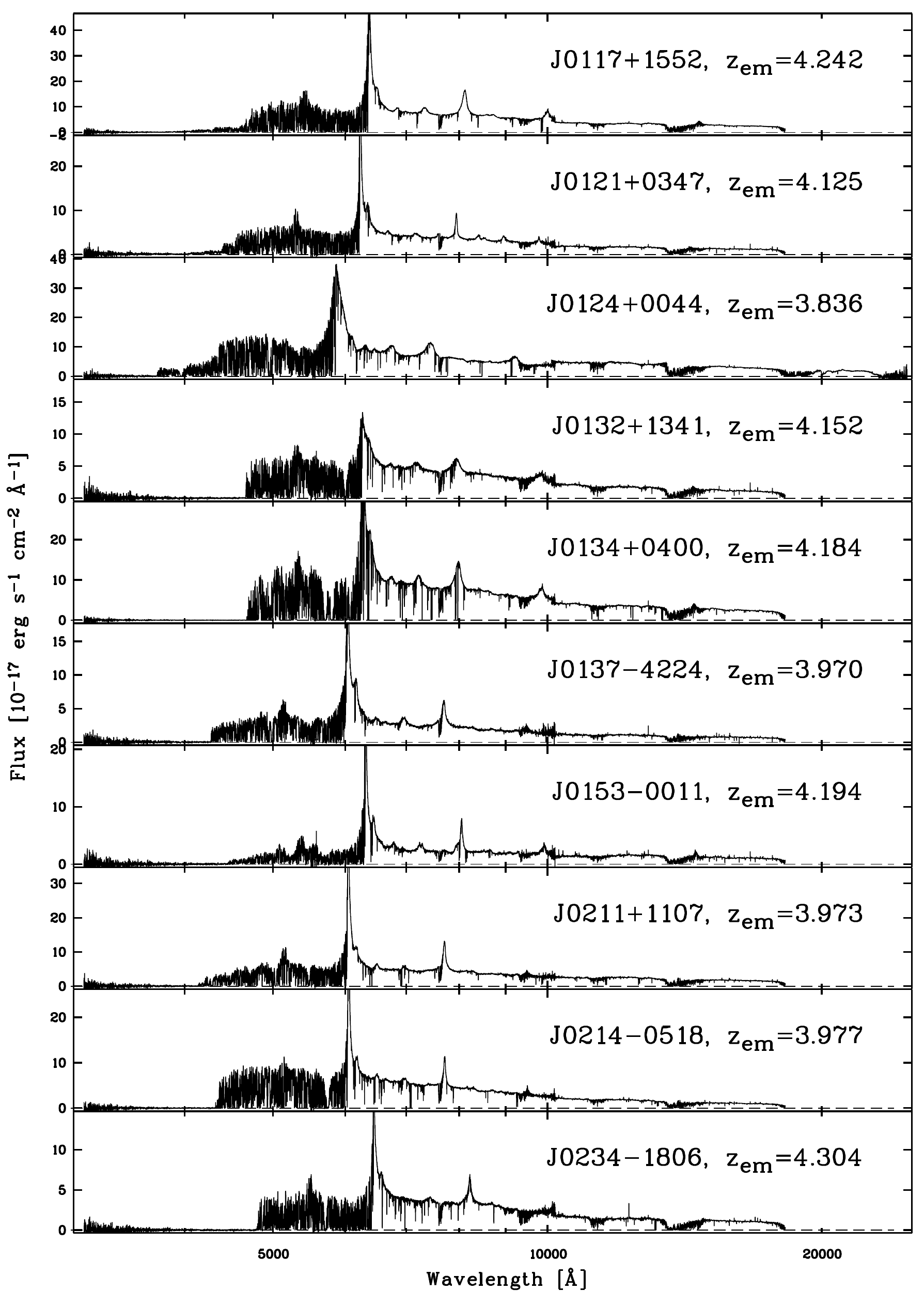}
\caption{Same as Fig.~\ref{fig_all}.}
\end{figure*}

\begin{figure*}
\includegraphics[width=170mm,clip]{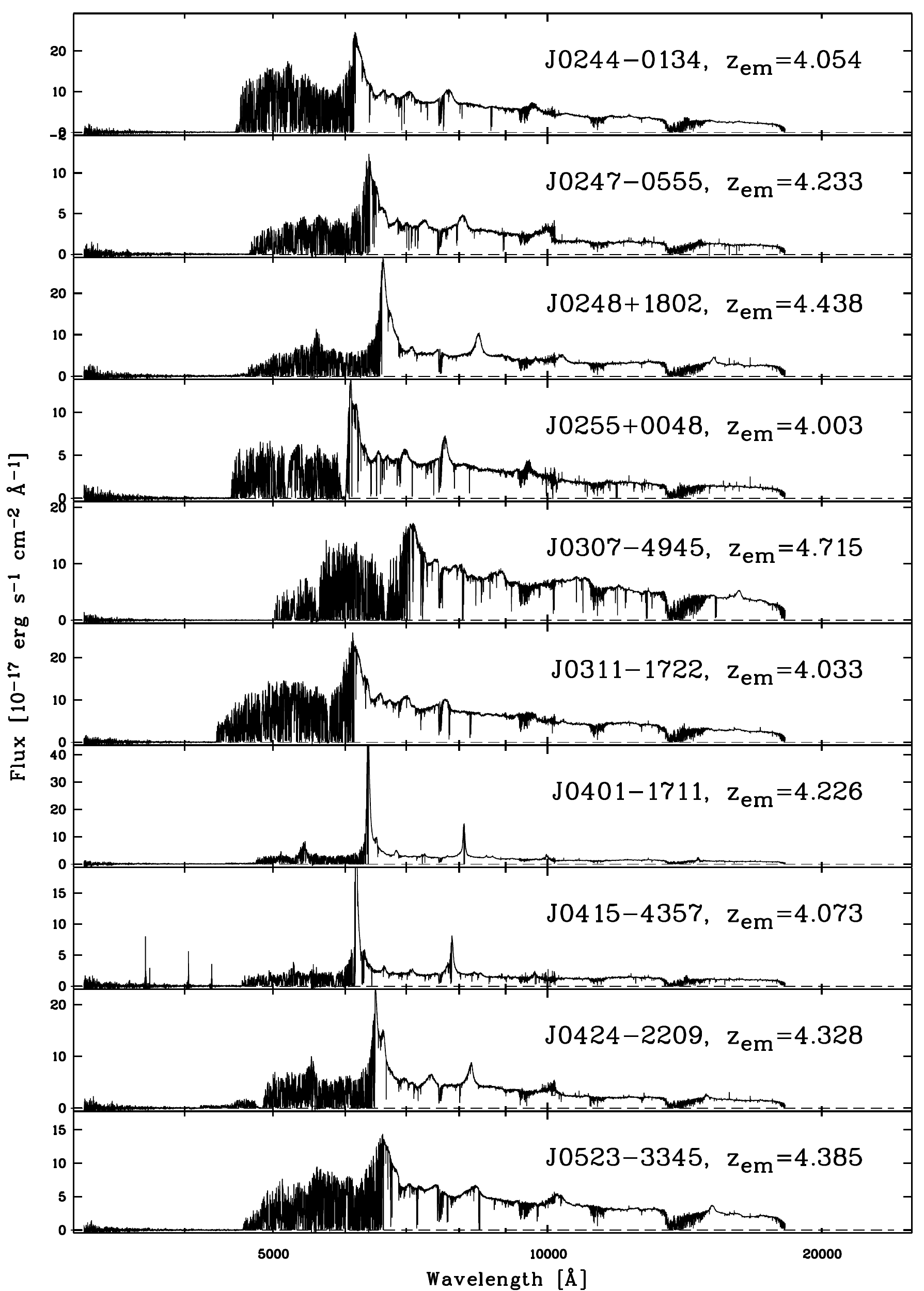}
\caption{Same as Fig.~\ref{fig_all}.}
\end{figure*}

\begin{figure*}
\includegraphics[width=170mm,clip]{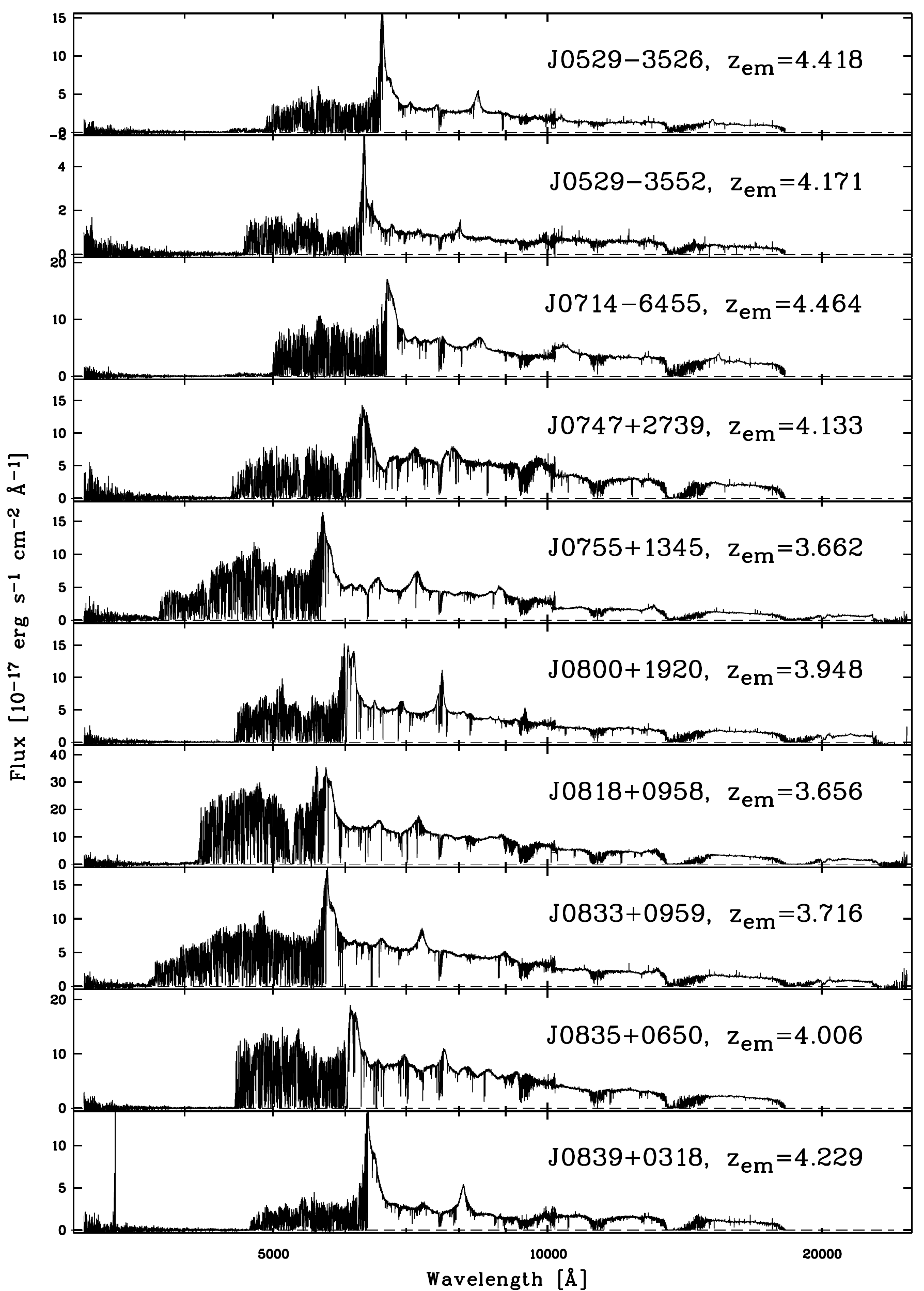}
\caption{Same as Fig.~\ref{fig_all}.}
\end{figure*}

\begin{figure*}
\includegraphics[width=170mm,clip]{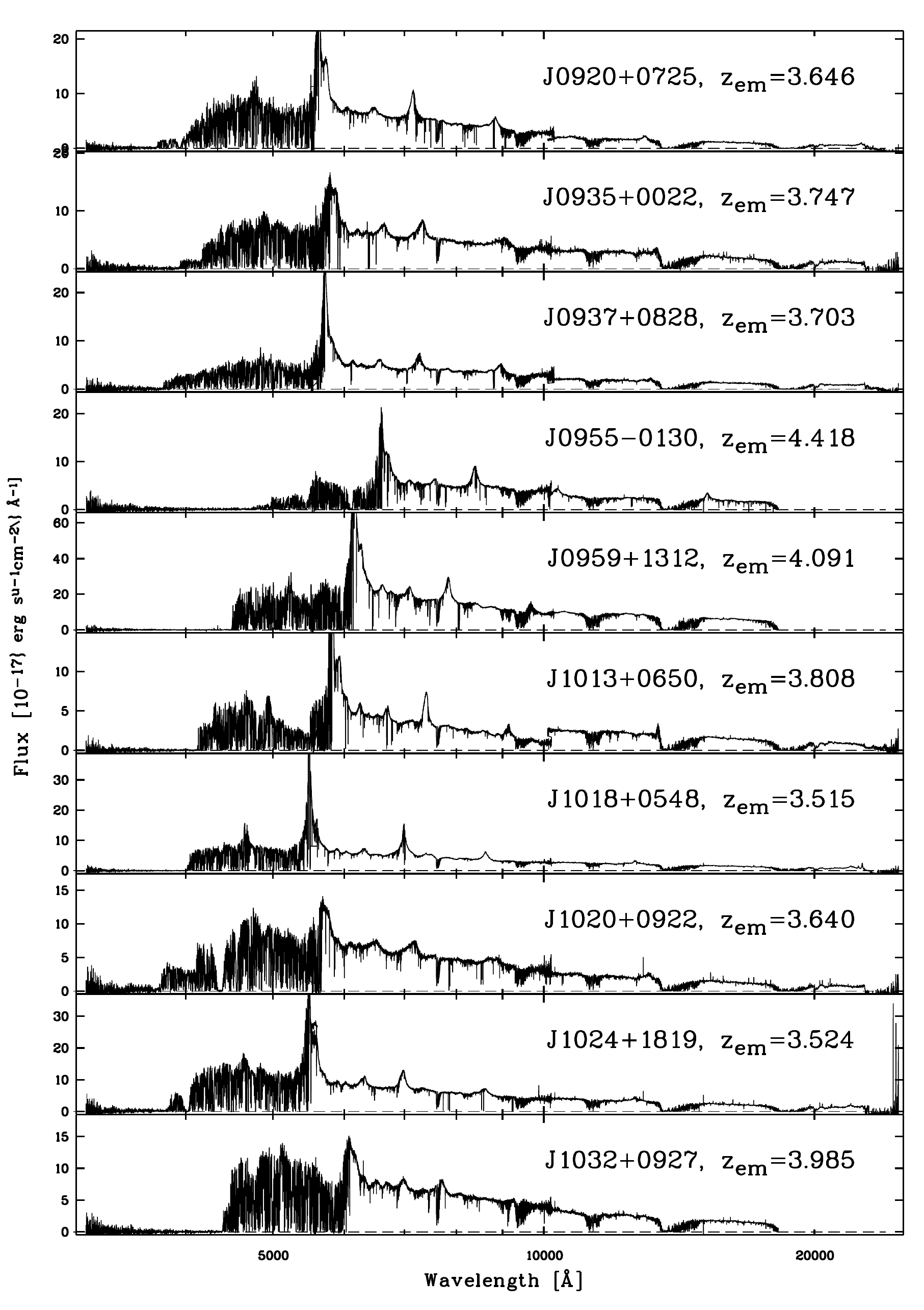}
\caption{Same as Fig.~\ref{fig_all}.}
\end{figure*}

\begin{figure*}
\includegraphics[width=170mm,clip]{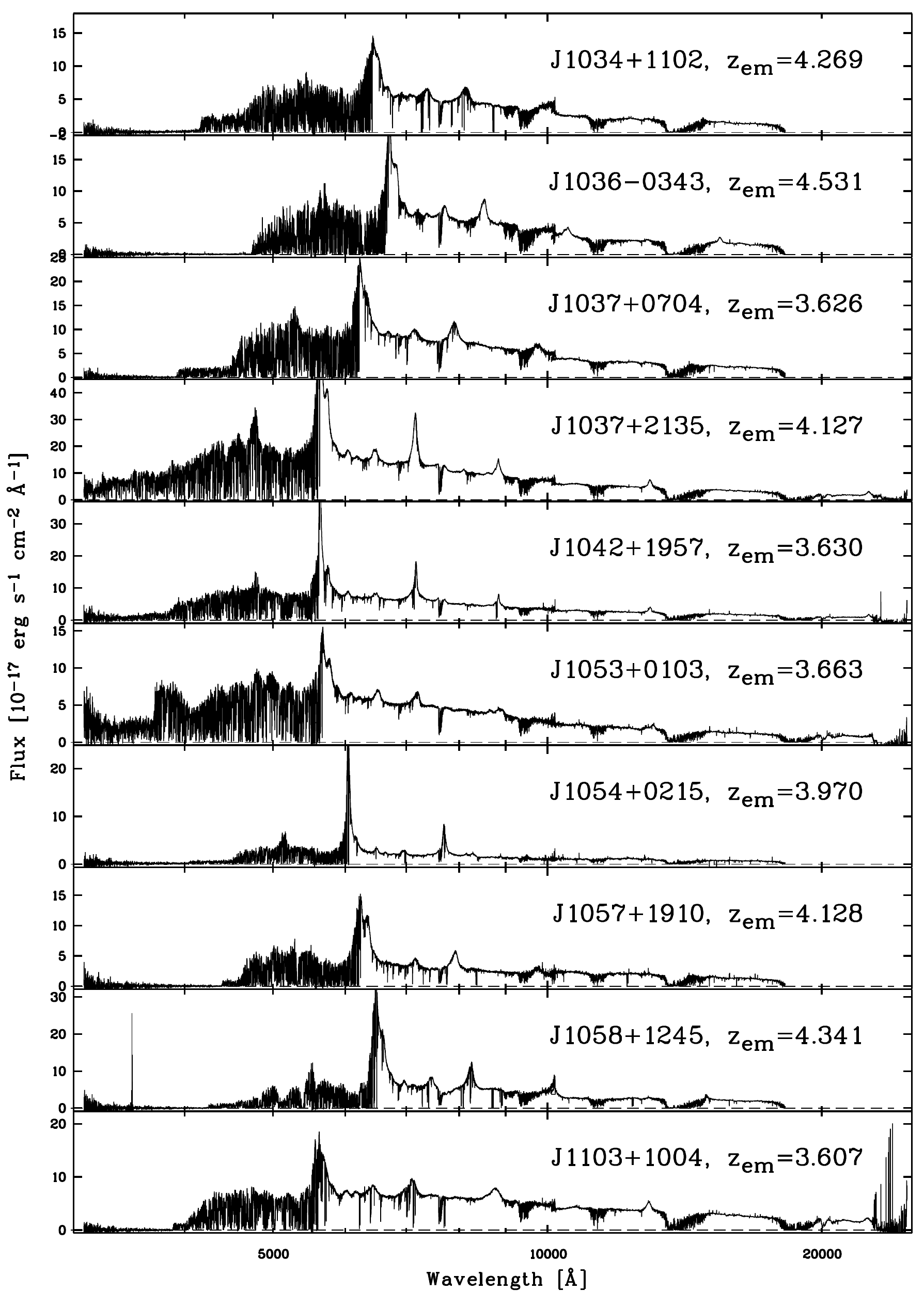}
\caption{Same as Fig.~\ref{fig_all}.}
\end{figure*}

\begin{figure*}
\includegraphics[width=170mm,clip]{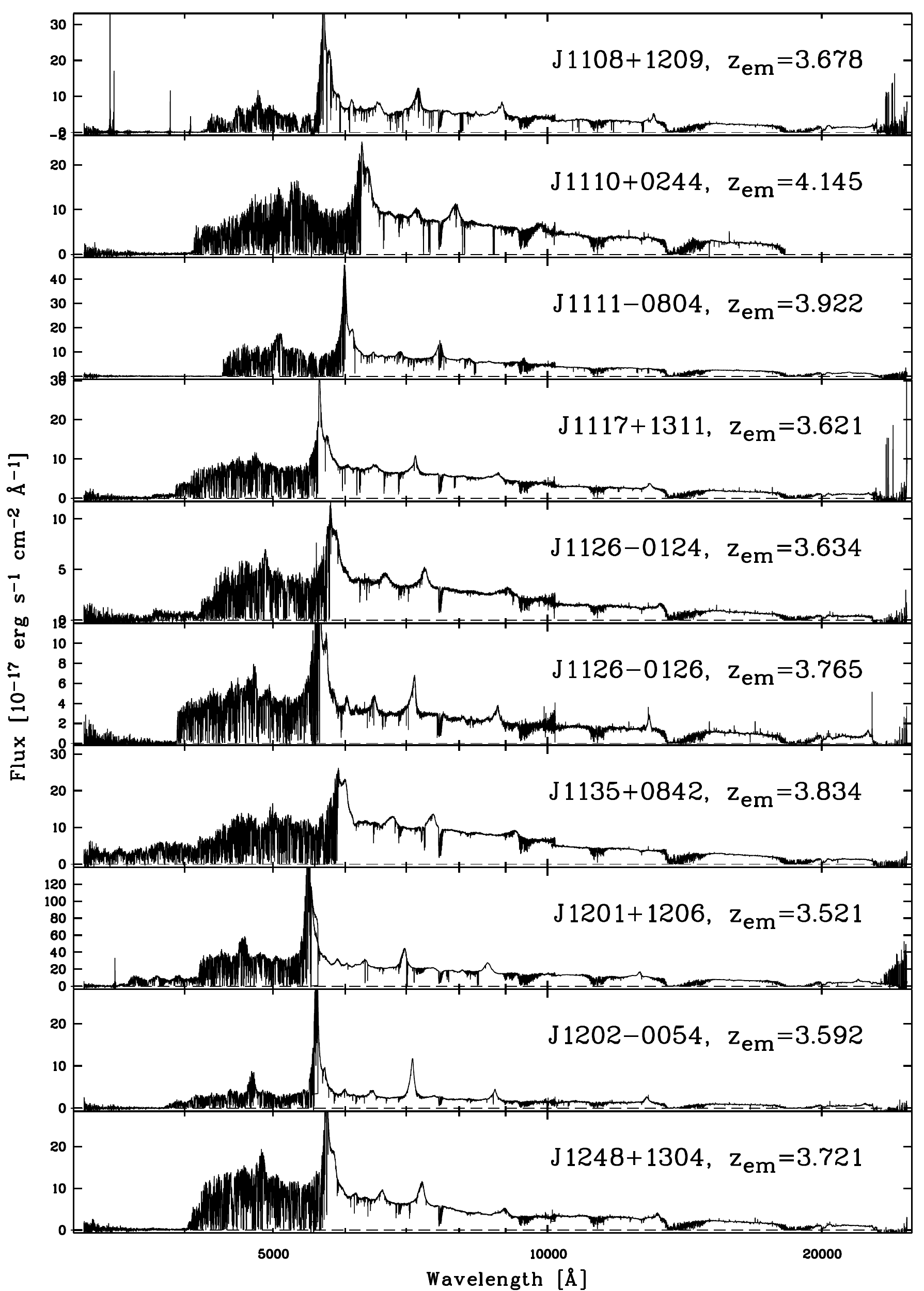}
\caption{Same as Fig.~\ref{fig_all}.}
\end{figure*}

\begin{figure*}
\includegraphics[width=170mm,clip]{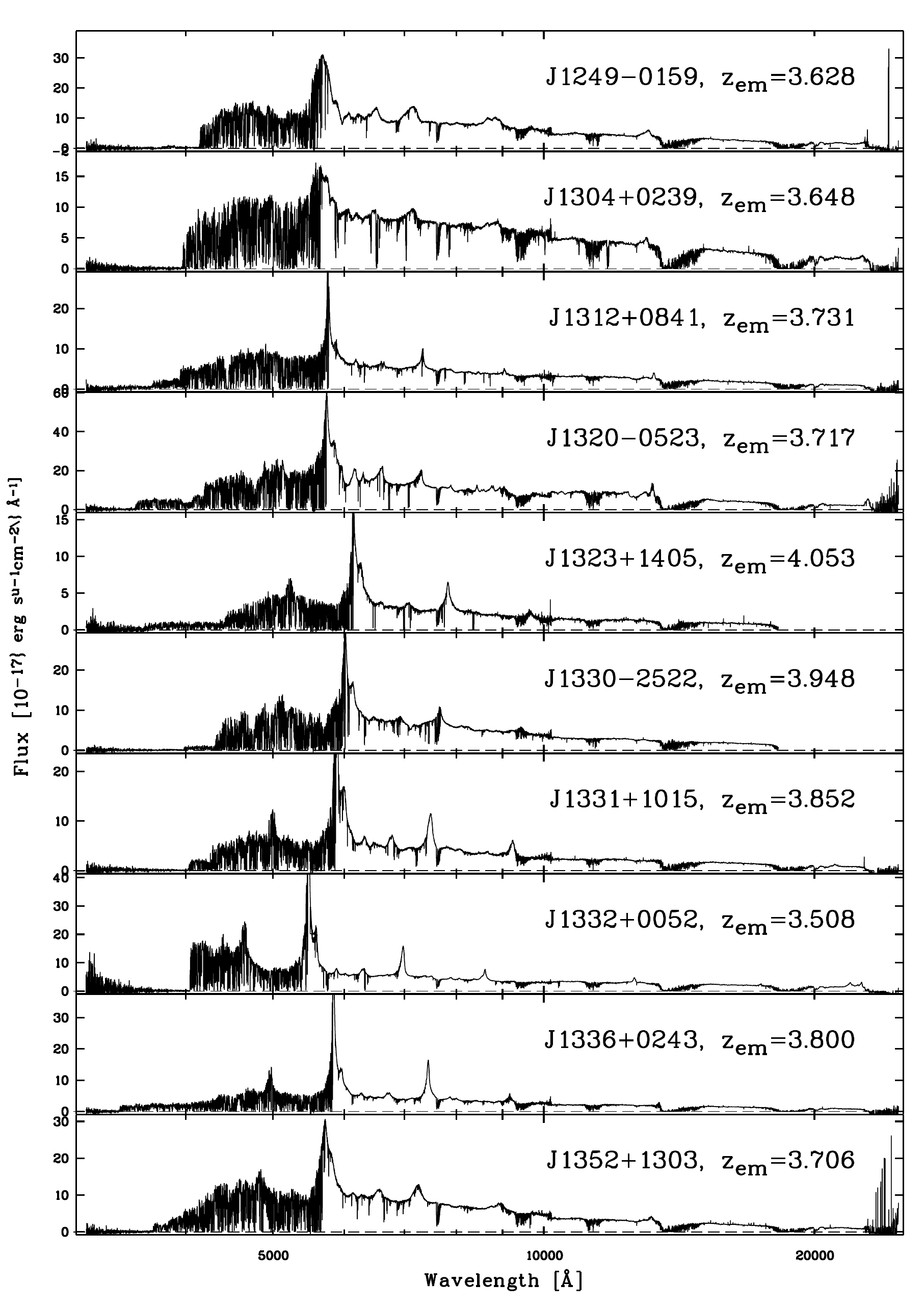}
\caption{Same as Fig.~\ref{fig_all}.}
\end{figure*}

\begin{figure*}
\includegraphics[width=170mm,clip]{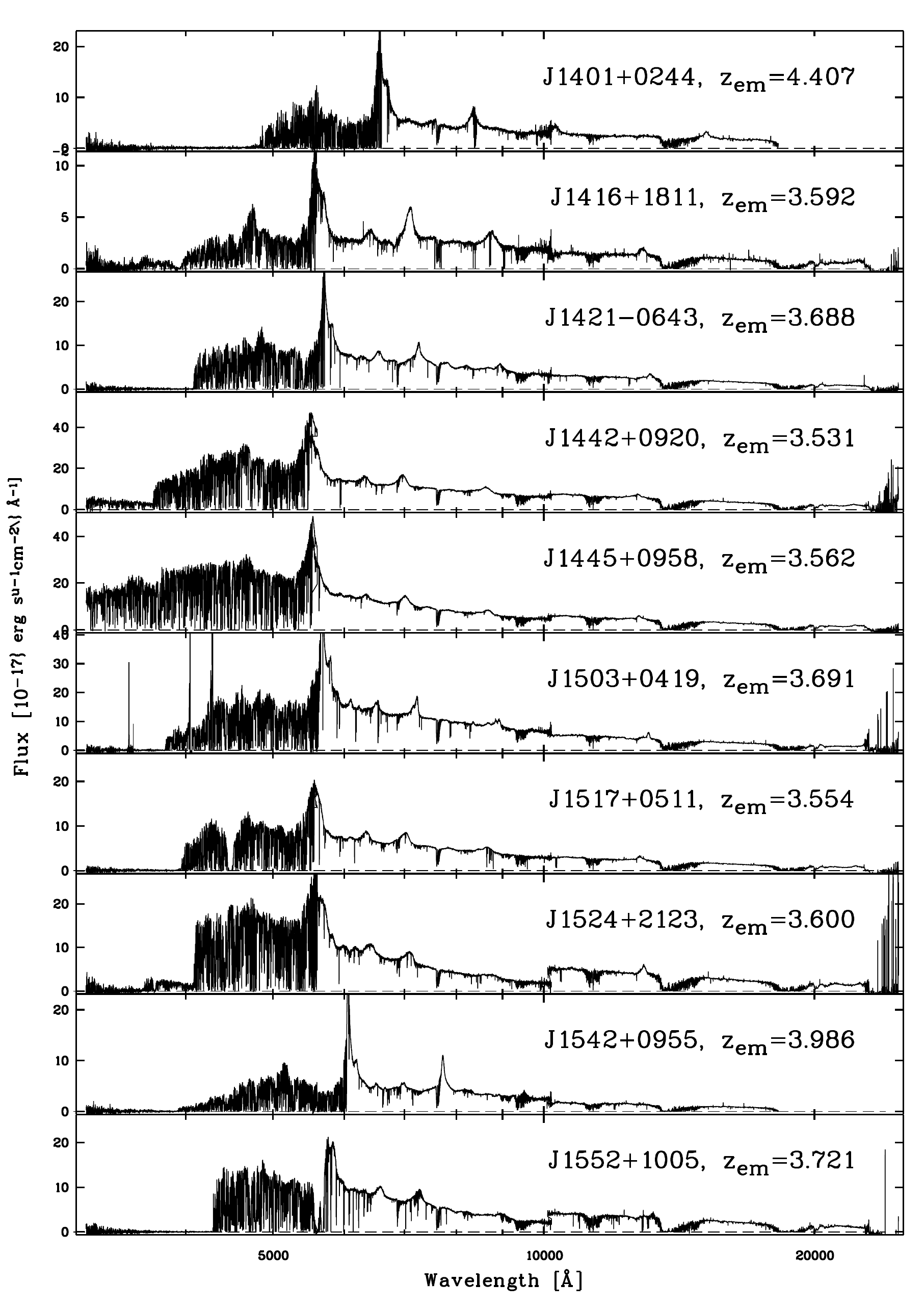}
\caption{Same as Fig.~\ref{fig_all}.}
\end{figure*}

\begin{figure*}
\includegraphics[width=170mm,clip]{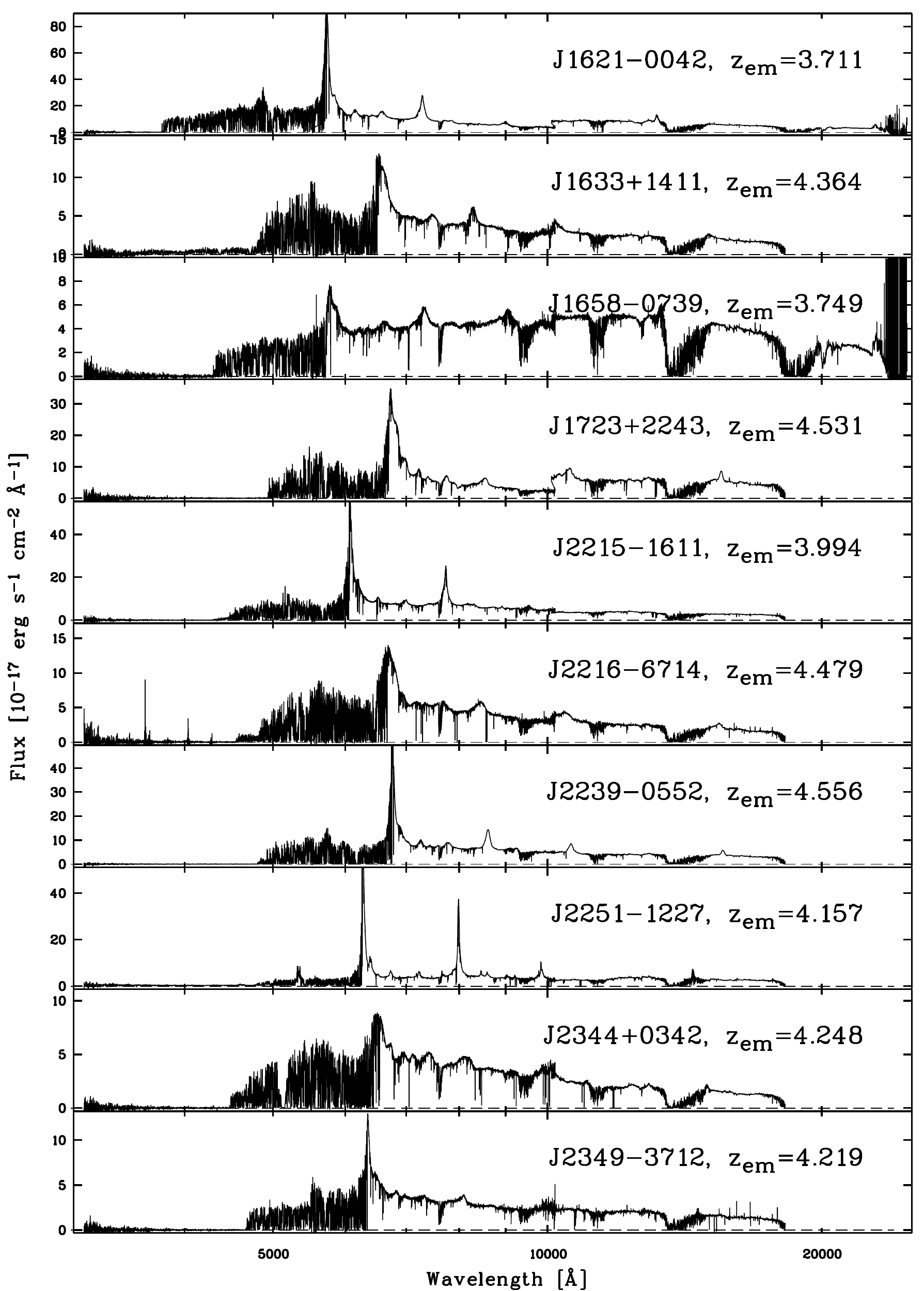}
\caption{Same as Fig.~\ref{fig_all}.
\label{fig_all_last}}
\end{figure*}

\label{lastpage}


\begin{thebibliography}{199}

\bibitem[Aguirre et al. (2004)]{aguirre2004}
Aguirre, A., Schaye, J., Kim, T.-S., Theuns, T., Rauch, M. \& 
Sargent, W. L. W.  2004 ApJ 602 38

\bibitem[Alam et al. (2015)]{DR12}
Alam, S., Albareti, F. D., Allende Prieto, C., et al. 2015, ArXiv:1501.00963

\bibitem[Becker, Rauch \& Sargent (2007)]{becker2007}
Becker, G. D., Rauch, M. \&  Sargent, W. L. W. 2007, ApJ, 662, 72

\bibitem[Becker, Rauch \& Sargent (2009)]{becker2009}
Becker, G. D., Rauch, M. \&  Sargent, W. L. W. 2009, ApJ, 698, 1010

\bibitem[Becker et al. (2012)]{becker2012}
Becker, G. D., Sargent, W. L. W., Rauch, M., \& Carswell, R. F. 2012, ApJ,
744, 91

\bibitem[Becker et al. (2015)]{becker2015}
Becker, G. D., Bolton, J. S., Madau, P., Pettini, M., Ryan-Weber,
E. V. \& Venemans, B. P. 2015, MNRAS, 447, 3402


\bibitem[Becker, White \& Helfand (1995)]{becker1995} Becker, R. H., White,
  R. L., \& Helfand, D. J. 1995, ApJ, 450, 559

\bibitem[Berg et al. (2016)]{berg2016}
Berg, T., Ellison, S. L., S\'anchez-Ram\'irez, R. et al. et al. 2016, MNRAS,
submitted. 

\bibitem[Bergeron et al. (2004)]{bergeron2004}
Bergeron J. et al. 2004, The Messenger, 118, 40

\bibitem[Brunner et al. (2002)]{brunner2002} Brunner, R., Djorgovski, S. G.,
  Prince, T.  \& Szalay, A., “Massive Data Sets in Astronomy”, in Handbook of
  Massive Data Sets, eds. J. Abello et al., Dordrecht: Kluwer Academic Publ.,
  2002, pp. 931-979.



\bibitem[Calverley et al. (2011)]{calverley2011} Calverley, A. P., Becker,
  G. D., Haehnelt, M. G. \& Bolton, J.  S.  2011, MNRAS, 412, 2543

\bibitem[Capellupo et al. (2015)]{capellupo2015}
	Capellupo, D. M., Netzer, H., Lira, P., Trakhtenbrot, B. \&
        Mejía-Restrepo, J.  2015, MNRAS. 446. 3427


\bibitem[Chen et al. (2010)]{chen2010}
	Chen, H.-W., Wild, V., Tinker, J. L., Gauthier,
        J.-R., Helsby, J. E., Shectman, S. A. \& Thompson, I. 
        B.  2010,  ApJ 724, 176

\bibitem[Croft et al. (2002)]{croft2002}	
	Croft, R. A. C., Weinberg, D. H., Bolte, M., Burles, S., 
        Hernquist, L., Katz, N., Kirkman, D. \& Tytler, D. 	
	2002 ApJ, 581, 20 


\bibitem[Croft et al. (1998)]{croft1998} Croft, R. A. C., Weinberg, D. H.,
  Katz, N. \& Hernquist, L. 1998, ApJ, 495, 44

\bibitem[Cutri et al. (2003)]{cutri2003} Cutri, R. M., Skrutskie, M. F., van
  Dyk, S., et al. 2003, 2MASS All Sky Catalog of point sources.

\bibitem[Dall'Aglio, Wisotzki \& Worseck (2008)]{dallaglio2008} Dall'Aglio,
  A., Wisotzki, L. \& Worseck, G.  2008, A\&A, 491,465


\bibitem[De Rosa et al. (2014)]{derosa2014} De Rosa, G., Venemans, B. P.,
  Decarli, R., Gennaro, M., Simcoe, R. A., Dietrich, M., Peterson, B. M.,
  Walter, F., Frank, S., McMahon, R. G., Hewett, P. C., Mortlock, D. J. \&
  Simpson, Chris 2014, ApJ, 790, 145

\bibitem[Djorgovski (2005)]{djorgovski2005}
Djorgovski, S. G., “Virtual Astronomy, Information
Technology, and the New Scientific Methodology”, in IEEE
Proc. of CAMP05: Computer Architectures for Machine
Perception, eds. V. Di Gesu \& D. Tegolo, 2005, p. 125.



\bibitem[{D'Odorico} {et~al.}(2004)]{dodorico2004} {D'Odorico}, V., Cristiani,
  S., Romano, D., Granato, G. L. \& Danese, L. 2004, MNRAS, 351, 976


\bibitem[{D'Odorico} {et~al.}(2008)]{dodorico2008} {D'Odorico}, V.,
  {Bruscoli}, M., {Saitta}, F., {Fontanot}, F., {Viel}, M., {Cristiani}, S.,
  \& {Monaco}, P. 2008, MNRAS, 389, 1727

\bibitem[{D'Odorico} {et~al.}(2010)]{dodorico2010} {D'Odorico}, V., Calura,
  F., Cristiani, S., \& Viel, M. 2010, MNRAS, 401, 2715


\bibitem[D'Odorico et al. (2013)]{dodorico2013}
	D'Odorico, V., Cupani, G., Cristiani, S., Maiolino, R., Molaro, P.,
        Nonino, M., Centurión, M., Cimatti, A., di Serego Alighieri, S.,
        Fiore, F., Fontana, A., Gallerani, S., Giallongo, E., Mannucci, F.,
        Marconi, A., Pentericci, L., Viel, M. \& Vladilo, G. 2013, MNRAS, 435,
        1198

\bibitem[Dietrich et al.(2002)]{dietrich2002} Dietrich, M., 
Appenzeller, I., Vestergaard, M., \& Wagner, S.~J.\ 2002, \apj, 564, 581 

\bibitem[Dietrich et al.(2003)]{dietrich2003}
Dietrich, M., Hamann, F., Appenzeller, I. \& Vestergaard, M.,
 2003 ApJ 596, 817

\bibitem[Dietrich et al. (2009)]{dietrich2009} Dietrich, M., et al., 2009, ApJ,
  696, 1998  


\bibitem[Eisenstein et al. (2011)]{eisenstein2011}
Eisenstein, D. J., Weinberg, D. H., Agol, E., et al. 2011, AJ, 142, 72

\bibitem[Flesch (2015)]{flesch2015}
Flesch, E. 2015, PASA, 32, 10

\bibitem[Freudling et al. (2013)]{freudling2013}
Freudling, W., Romaniello, M., Bramich, D. M., Ballester, P., Forchi, V.,
García-Dabl\'o, C. E., Moehler, S. \& Neeser, M. J. 2013, A\&A, 559, 96 

\bibitem[Hamann & Ferland (1999)]{hamann1999} Hamann, F. \& Ferland, G. 1999,
  ARA\&A, 37, 487

\bibitem[Hamann et al. (2002)]{hamann2002} Hamann, F., Korista, K. T.,
  Ferland, G. J., Warner, C. \& Baldwin, J. 2002, ApJ, 564, 592



\bibitem[Ho et al. (2012)]{ho2012} Ho, L. C., Goldoni,
  P., Dong, X., Greene, J. E. \& Ponti, G. 2012,  ApJ, 754, 11 

\bibitem[Horne(1986)]{horne1986} Horne, K. 1986, PASP, 98, 609 

\bibitem[Ir\v{s}i\v{c} et al. (2016)]{irsic2016} Ir\v{s}i\v{c}, V., Viel, M.,
  Berg, T. A. M.,  et al., MNRAS, submitted. 

\bibitem[Irwin, McMahon \& Hazard (1991)]{irwin1991} 
  Irwin, M., McMahon, R. G. \& Hazard, C. 1991, ASPC, 21, 117

\bibitem[Jiang et al. (2007)]{jiang2007} 
Jiang, L., Fan, X., Vestergaard, M., Kurk, J. D., Walter,
F., Kelly, B. C. \& Strauss, M. A.  2007, AJ, 134, 1150

\bibitem[Jones et al. (2013)]{jones2013} 
Jones, A., Noll, S., Kausch, W., Szyszka, C. \&  Kimeswenger, S. 2013, A\&A,
560, 91

\bibitem[Kelson(2003)]{kelson2003} Kelson, D. D.  2003, PASP, 115,
  688  


\bibitem[Kim et al.(2002)]{kim2002}
Kim, T.-S., Carswell, R. F., Cristiani, S., D'Odorico, S., Giallongo, E. 
	2002 MNRAS, 335, 555

\bibitem[Kirkman et al.(2003)]{kirkman2003}
Kirkman, D., Tytler, D., Suzuki, N., O'Meara, J. M. \& Lubin, D.  2003
ApJS, 149, 1

\bibitem[Kirkman et al.(2005)]{kirkman2005}
Kirkman, D., Tytler, D., Suzuki, N., Melis, C.,  
	Hollywood, S., James, K., So, G., Lubin, D.,  
	Jena, T., Norman, M.~L. \& Paschos, P. 2005, MNRAS, 360, 373

\bibitem[Kondo et al.(2008)]{Kondo2008} Kondo, S., et al. 2008, in
  Astronomical Society of the Pacific Conference Series, Vol. 399, Panoramic
  Views of Galaxy Formation and Evolution, ed. T. Kodama, T. Yamada, \&
  K. Aoki, 209

\bibitem[Ledoux et al. (2003)]{ledoux2003}
	Ledoux, C., Petitjean, P. \& Srianand, R. 2003 MNRAS, 346, 209 



\bibitem[Lu et al.(1996)]{lu1996}
Lu, L., Sargent, W. L. W., Barlow, T. A., Churchill, C. W. \& Vogt, S. S. 	
	1996 ApJS,  107, 475


\bibitem[Marziani et al.(2009)]{marziani2009}
Marziani, P., Sulentic, J. W., Stirpe, G. M., Zamfir, S. \& Calvani, M.
2009, A\&A, 495, 83

\bibitem[Matejek \& Simcoe(2012)]{Matejek2012} Matejek, M.~S., \&
  Simcoe, R.~A.\ 2012, ApJ, 761, 112

\bibitem[M{\'e}nard et al.(2011)]{2011MNRAS.417..801M} M{\'e}nard, B., 
Wild, V., Nestor, D., et al.\ 2011, MNRAS, 417, 801 


\bibitem[Molaro et al.(2013)]{molaro2013} Molaro, P., Centurión, M., Whitmore,
  J. B., Evans, T. M., Murphy, M. T., Agafonova, I. I., Bonifacio, P.,
  D'Odorico, S., Levshakov, S. A., L\'opez, S., Martins, C. J. A. P., Petitjean,
  P., Rahmani, H., Reimers, D., Srianand, R., Vladilo, G. \& Wendt, M. 2013
  A\&A, 555, 68

\bibitem[Murphy, Webb \& Flambaum (2003)]{murphy2003} Murphy, M. T., Webb,
  J. K. \& Flambaum, V. V. 2003, MNRAS, 345, 609




\bibitem[Noll et al.(2012)]{noll2012}
Noll, S., Kausch, W., Barden, M., Jones, A. M., Szyszka, C., Kimeswenger,
S. \& Vinther, J.  2012 A\&A, 543, 92


\bibitem[Noterdaeme et al. (2012a)]{noterdaeme2012a} 
Noterdaeme, P., Laursen, P., Petitjean, P., Vergani, S. D., Maureira, M. J.,
Ledoux, C., Fynbo, J. P. U., López, S. \&  Srianand, R.  2012a A\&A 540, 63  

\bibitem[Noterdaeme et al. (2012b)]{noterdaeme2012b} Noterdaeme, P., Petitjean,
  P., Carithers, W. C., Pâris, I., Font-Ribera, A., Bailey, S., Aubourg, E.,
  Bizyaev, D., Ebelke, G., Finley, H., Ge, J., Malanushenko, E., Malanushenko,
  V., Miralda-Escud\'e, J., Myers, A. D., Oravetz, D., Pan, K., Pieri, M. M.,
  Ross, N. P., Schneider, D. P., Simmons, A. \& York, D. G. 2012b, A\&A, 547,
  L1

\bibitem[O'Meara et al. (2015)]{omeara2015} O'Meara, J. M., Lehner, N., Howk,
  J. C., Prochaska, J. X., Fox, A. J., Swain, M. A., Gelino, C. R., Berriman,
  G. B. \& Tran, H.  2015, AJ, 150, 111

\bibitem[Palanque-Delabrouille et al. (2013)]{palanque2013}
  Palanque-Delabrouille, N. et al. 2013, A\&A, 559, 85



\bibitem[P\^aris et al. (2012)]{paris2012}
Pâris, I., Petitjean, P., Aubourg, É., et al. 2012, A\&A, 548, A66

\bibitem[P\^aris et al. (2014)]{paris2014}
Pâris, I., Petitjean, P., Aubourg, É., et al. 2014, A\&A, 563, 54


\bibitem[Patat \& Hussain (2013)]{patat2013}
Patat, F. \& Hussain, G. A. J. 2013, in Organizations,
People and Strategies in Astronomy 2 (OPSA 2),
ed. Heck, A., 231


\bibitem[Peroux et al. (2011)]{peroux2011}
Péroux, C., Bouch\'e, N. Kulkarni, V. P., York, D. G. \&  Vladilo, G. 2011
MNRAS, 410, 2237  


\bibitem[Perrotta et al. (2016)]{perrotta2016}
Perrotta, S., 
D'Odorico, V.,  
Prochaska, J. X., 
Cristiani, S.,
Cupani, G., 
Ellison, S. L., 
L\'opez, S., 
Becker, G. D., 
Berg, T. A. M., 
Christensen, L., 
Denney, K. D., 
Hamann, F., 
P\^aris, I., 
Vestergaard, M. \&
Worseck, G. 2016, MNRAS, in press. 


\bibitem[Prochaska et al. (2003)]{prochaska2003}
	Prochaska, J. X., Gawiser, E., Wolfe, A. M., Castro, S. \& 
        Djorgovski, S. G. 2003 ApJ, 595, 9 

\bibitem[Prochaska, Worseck \& O'Meara (2009)]{prochaska2009}
	Prochaska, J. X., Worseck, G. \& O'Meara, J. M. 
	2009, ApJ, 705, 113

\bibitem[Prochaska, O'Meara \& Worseck (2010)]{prochaska2010}	
	Prochaska, J. X., O'Meara, J. M. \& Worseck, G. 2010, ApJ, 718, 392 



\bibitem[Prochaska \& Wolfe (2009)]{prochaska2009a}
	Prochaska, J. Xavier; Wolfe, Arthur M. 2009, ApJ, 696, 1543

\bibitem[Rafelski et al. (2013)]{rafelski2012} Rafelski, M. Wolfe, A. M.,
  Prochaska, J. X., Neeleman, M. \& Mendez, A. J.  2012, ApJ, 755, 89

\bibitem[Rudie et al. (2012)]{rudie2012} Rudie, G. C., Steidel, C. C.,
  Trainor, R. F., Rakic, O., Bogosavljevi\'c, M., Pettini, M., Reddy, N.,
  Shapley, A. E., Erb, D. K. \& Law, D. R. 2012 ApJ, 750, 67


\bibitem[Ryan-Weber et al. (2009)]{ryan2009}
Ryan-Weber, E. V., Pettini, M., Madau, P. \& Zych, B. J. 2009, MNRAS, 395,
1476 

\bibitem[S\'anchez-Ram\'irez et al. (2016)]{sanchez2015}
S\'anchez-Ram\'irez,  Ellison, S. L., Prochaska, J. X., Berg, T. A. M.,
L\'opez, S., D'Odorico, V.,  
 Becker, G. D., Christensen, L., Cupani, G., Denney, K. D., P\^aris, I., 
 Worseck, G. \& Gorosabel, J. 2016, MNRAS, 456, 4488




\bibitem[Scannapieco et al. (2006)]{scannapieco2006} Scannapieco, E., Pichon,
  C., Aracil, B., Petitjean, P., Thacker, R. J., Pogosyan, D., Bergeron, J. \&
  Couchman, H. M. P.  2006, MNRAS, 365, 615

\bibitem[Schaye et al. (2000)]{schaye2000}
Schaye, J., Theuns, T., Rauch, M., Efstathiou, G. \&  Sargent, W. L. W. 
2000, MNRAS, 318, 817 

\bibitem[Schneider et al. (2010)]{schneider2010}
Schneider, D. P., Richards, G. T., Hall, P. B., et al. 2010, AJ, 139, 2360

\bibitem[Simcoe et al.(2011)]{Simcoe2011}  Simcoe, R. A., Cooksey, K. L.,
  Matejek, M., Burgasser, A. J., Bochanski, J., Lovegrove, E., Bernstein,
  R. A., Pipher, J. L., Forrest, W. J., McMurtry, C., Fan, X. \& O'Meara,
  J. 2011, ApJ, 743, 21

\bibitem[Songaila (2005)]{songaila2005} Songaila, A. 2005 AJ, 130, 1996

\bibitem[Songaila \& Cowie.(2010)]{songaila2010} Songaila, A. \& Cowie,
  L. L. 2010, ApJ, 721, 1448

\bibitem[Srianand et al. (2004)]{srianand2004}
Srianand, R., Chand, H., Petitjean, P. \& Aracil, B. 2004, PhRvL,
92, 121302

\bibitem[Storrie-Lombardi et al. (1994)]{storrie1994} Storrie-Lombardi, L. J.,
  McMahon, R. G., Irwin, M. J. \& Hazard, C. 1994, ApJ, 427, 13


\bibitem[Sulentic et al.(2006)]{sulenticetal06} Sulentic, J.~W., Repetto, P.,
  Stirpe, G.~M., Marziani, P., Dultzin-Hacyan, D., \& Calvani, M. 2006, AAp,
  456, 929 (Paper II) %

\bibitem[Sulentic et al.(2004)]{sulenticetal04} Sulentic, J.~W., Stirpe,
  G.~M., Marziani, P., Zamanov, R., Calvani, M., \& Braito, V.\ 2004, AAp,
  423, 121 (Paper I) %

\bibitem[Suzuki et al.(2005)]{suzuki2005} {{Suzuki}, N. and {Tytler}, D. and
  {Kirkman}, D. and {O'Meara}, J.~M. and {Lubin}, D.}, 2005, ApJ, 618, 592

\bibitem[Vernet et al. (2011)]{Vernet2011} Vernet et al. 2011, A\&A, 536A, 105

\bibitem[Vestergaard \& Peterson (2006)]{vestergaard2006} Vestergaard, M. \&
  Peterson, B.M. 2006, ApJ, 641, 689 

\bibitem[Vestergaard \& Osmer (2009)]{vestergaard2009} Vestergaard, M. \&
  Osmer, P.S. 2009, ApJ, 699, 800 


\bibitem[Viel et al. (2004)]{viel2004} 
Viel, M., Haehnelt, M. G., \&  Springel, V. 2004, MNRAS, 354, 684 

\bibitem[Viel et al. (2009)]{viel2009} Viel, M., Bolton, J. S., \& Haehnelt,
  M. G. 2009, MNRAS, 399, L39 

\bibitem[Viel et al. (2013)]{viel2013} 
Viel, M., Becker, G. D., Bolton, J. S. \& Haehnelt, M. G. 2013, PhRvD,
88, 043502   

\bibitem[Worseck \& Prochaska (2011)]{Worseck2011} 
Worseck, G. \& Prochaska, J. X. 2011, ApJ, 728, 23

\bibitem[Worseck et al. (2014)]{worseck2014} 
Worseck, G., Prochaska, J. X., O'Meara, J. M., Becker, G. D.,
Ellison, S. L., L\'opez, S., Meiksin, A., Ménard, B., Murphy,
M. T. \&  Fumagalli, M. 2014 MNRAS, 445, 1745

\bibitem[Wright et al. (2010)]{wright2010}
Wright, E. L., Eisenhardt, P. R. M., Mainzer, A. K., et al. 2010, AJ, 140, 1868

\bibitem[Wolfe, Gawiser \& Prochaska (2005)]{wolfe2005}
Wolfe, A. M., Gawiser, E. \& Prochaska, J. X. 2005, ARA\&A, 43, 861 

\bibitem[York et al. (2000)]{york2000}
York, D. G., Adelman, J., Anderson, Jr., J. E., et al. 2000, AJ, 120, 1579

\bibitem[Zafar et al. (2013)]{zafar2013} 	Zafar, T., Popping, A.  \&
  Péroux, C.	2013,  A\&A, 556, A140 

\bibitem[Zhu \& M\'enard (2013)]{zhu2013} 	Zhu, G. \& M\'enard, B.	2013,
  ApJ, 770, 130 

\bibitem[Zuo et al. (2015)]{zuo2015}
Zuo, W., Wu, X-B., Fan, X., Green, R., Wang, R. \& Bian, F. 2015, ApJ, 799,
189 

\end{thebibliography}
\end{document}